\documentclass[twocolumn,epjc3]{svjour3}  

\usepackage{graphicx}  
\usepackage{amssymb}
\usepackage{amsmath}
\usepackage{color}  
\definecolor{darkgreen}{rgb}{0.2,0.5, 0.2}

\usepackage{dcolumn,booktabs}
\newcolumntype{d}[1]{D{.}{.}{#1}}
\newcommand\mc[1]{\multicolumn{1}{c}{#1}} 

\usepackage{lipsum}

\definecolor{mbscolor}{rgb}{0.60, 0.0, 0.65}

\usepackage{dsfont}

\journalname{Eur. Phys. J. A}

\begin{document}

\title{Skyrme-Hartree-Fock-Bogoliubov mass models on a 3D Mesh: Effect of triaxial shape}


\author{Guillaume Scamps\thanksref{e1,addr1}
        \and
        Stephane Goriely\thanksref{addr1} 
        \and
        Erik Olsen\thanksref{addr1}
        \and
        Michael Bender\thanksref{addr2}
        \and
        Wouter Ryssens\thanksref{addr1,addr3}        
}
\thankstext{e1}{e-mail: guillaume.scamps@ulb.be}

\institute{Institut d'Astronomie et d'Astrophysique, Universit\'e Libre de Bruxelles, Campus de la Plaine CP 226, 1050 Brussels, Belgium \label{addr1}
           \and
           Universit{\'e} de Lyon, Universit{\'e} Claude Bernard Lyon 1, CNRS, IP2I Lyon / IN2P3, UMR 5822, F-69622, Villeurbanne, France \label{addr2}
           \and
           Center for Theoretical Physics, Sloane Physics Laboratory, Yale University, New Haven, Connecticut 06520, USA \label{addr3}
}

\maketitle

\begin{abstract}
The modeling of nuclear reactions and radioactive decays in astrophysical or earth-based conditions
requires detailed knowledge of the masses of essentially all nuclei. 
Microscopic mass models based on nuclear energy density functionals (EDFs) can 
be descriptive and used to provide this information. The concept of intrinsic symmetry breaking 
is central to the predictive power of EDF approaches, yet is generally not 
exploited to the utmost by mass models because of the computational demands 
of adjusting up to about two dozen parameters to thousands of nuclear masses. 
We report on a first step to bridge the gap between what is presently feasible 
for studies of individual nuclei and large-scale models: we present 
a new Skyrme-EDF-based model that was adjusted using a three-dimensional 
coordinate-space representation, for the first time allowing for both axial and triaxial deformations during the adjustment process. To compensate for the substantial increase in computational cost brought by the latter, we have employed a committee of multilayer neural networks 
to model the objective function in parameter space and guide us towards the 
overall best fit. The resulting mass model BSkG1 is computed with the EDF model independently of the neural network. It yields a root mean square (rms) 
deviation on the 2457 known masses of 741 keV and an rms deviation on the 884 
measured charge radii of 0.024 fm.
\end{abstract}


\section{Introduction}
\label{Sec:intro} 
The study of nuclear structure properties enters all chapters of nuclear physics
and plays a key role not only in our understanding of fundamental nuclear theory
but also in nuclear applications, such as astronuclear physics \cite{Arn20}.
In particular, the properties of atomic nuclei directly impact the description 
of nuclear reactions which often concern exotic species for 
which no experimental data exist. Despite significant efforts over several decades, experimental information only covers
a fraction of the entire data set required for nuclear applications.
Neutron-rich nuclei are of particular interest, especially for understanding how
heavy elements are made through the rapid neutron capture process, or r-process,
\cite{Arn20,Arn07} and for exploring the limits of nuclear stability 
\cite{Erler_12,Wang_15,Neufcourt_20b}. 
A major challenge for nuclear theory is to make reliable extrapolations 
into regions beyond current experimental reach. To make such predictions, 
models should strive to (i) contain as many physical ingredients as feasible and
(ii) reproduce as accurately as possible known experimental data on relevant 
observables. The tool of choice for this endeavor is nuclear Density Functional Theory (DFT)~\cite{Ben03}, the key ingredient of which is the nuclear energy density 
functional (EDF), which represents an effective interaction based
on one-body densities and currents 
\cite{Klupfel_09,Baldo13,Erler_13,Kortelainen14,Goriely16,Goriely09,Bennaceur_20}. This 
tool allows for predictions across the entire nuclear chart, firmly founded on
a microscopic description of the nucleus.

Particularly important for applications is the \emph{global} reproduction of 
nuclear masses. The HFB-series based on the Brussels-Montr{\'e}al (BSk)
interactions\footnote{Note the nomenclature of the mass model and the Skyrme
interaction: each HFB-$n$ mass model is associated with a Skyrme interaction 
BSk$n$. For the model we present here, we use the acronym BSkG1 to refer to both the mass model and the Skyrme interaction.}~\cite{Goriely16,Goriely09,Goriely14,Goriely13a} have 
demonstrated that a high accuracy on the known masses (\textit{i.e.}\ typically with
a root mean square (rms) deviation lower than 0.7~MeV) can be combined with
a global description of other properties of finite nuclei (radii, densities, 
fission barriers, \ldots) and properties of infinite matter as predicted by 
ab-initio calculations. Aside from nuclear charge radii and infinite nuclear
matter properties, the adjustment protocol of these interactions includes 
all known atomic masses~\cite{Wan17,Wan21}. Their inclusion 
is central to the success of these interactions but also renders the 
construction of new global models extremely demanding.

Depending on the nature of the observables and the types of nuclei considered, 
adjustments for even a limited number of systems can be extremely computationally demanding. Out of necessity, nuclear configurations 
considered during adjustment are typically highly restricted, both by the
symmetries imposed on them and the choices made to represent them numerically. Such
limitations impact both the generality and the numerical precision of the 
resulting model and do not reflect state-of-the-art EDF calculations, 
which can now routinely be carried out in large variational spaces with 
little to no symmetry restrictions~\cite{Ver19,Cwiok05} for a few nuclei at a time.

For the BSk models, the adjustment procedure relied on (i)
the imposition of axial symmetry and (ii) a numerical representation 
of the single-particle wavefunctions through an expansion in a limited number of
harmonic oscillator states (at least in the deformed case)~\cite{Goriely14,Goriely09,Goriely13a,Goriely16}.
We present here a new mass model, BSkG1, that is free of these limitations while
achieving a comparable accuracy on all known nuclear masses. First,  
we have moved to a three-dimensional representation that allows us to include, 
for the first time, triaxial deformations during the adjustment 
procedure. Second, we have adopted a coordinate-space representation in terms 
of a Lagrange mesh~\cite{Baye86}, which presents the advantage of excellent 
convergence in terms of the basis size~\cite{Ryssens15b,Arzhanov16}.
Some applications also require a coordinate-space representation, such 
as the description of fission processes by time-dependent mean-field methods
~\cite{Simenel18,Scamps18} or of the nuclear pasta phases in neutron star (NS) crusts 
or core-collapse supernovae~\cite{Magierski02,Gogelein07,Pais14,Schuetrumpf19}.

Moving from a two-dimensional to a three-dimensio\-nal representation
comes at a steep computational cost. A typical Skyrme-Hartree-Fock-Bogoliubov calculation for a nucleus, on a laptop, is of the order of a few seconds for a two-dimensional calculation, as opposed to several minutes for a three-dimensional one. To mitigate this, we have developed 
a new optimization procedure that is well suited to adjust large numbers of 
parameters to large data sets, employing machine learning tools inspired by 
Refs.~\cite{Bay18,Las20}. Such techniques are becoming ubiquitous in nuclear physics 
\cite{Las20,Sch15,Sch20,Niu18,Niu19,Bolla20,Neufcourt_18,Neufcourt_19,Neufcourt_20a}.

The paper is organized as follows. In Sec.~\ref{Sec:Ingredients}, we present the 
framework of the mass model. In Sec.~\ref{Sec:Fit}, a new method to adjust the 
parameters using machine learning techniques is proposed. The results of the 
BSkG1 mass model are discussed in Sec.~\ref{Sec:BSKG1}. Our conclusions and outlook are 
presented in Sec.~\ref{Sec:Summary}.

\section{Ingredients of the mass model}
\label{Sec:Ingredients} 

\subsection{The nuclear binding energy}
\label{Sec:Energy} 

 The mass model discussed here, like the BSk models before, describes the atomic 
nucleus by means of an auxiliary Hartree-Fock-Bogoliubov (HFB) many-body state 
$|\Phi \rangle$. For any such auxiliary state, we define the total 
binding energy $E_{\rm tot}$ of a nucleus as follows:
\begin{align}
E_{\rm tot}(|\Phi\rangle) &= E_{\rm HFB} + E_{\rm corr} \, , 
\label{eq:Etot}
\end{align}
where we refer to $E_{\rm HFB}$ as the self-consistent mean-field 
HFB energy and $E_{\rm corr}$ is a set of perturbative corrections. The 
distinction between both terms will be discussed in Sec.~\ref{sec:semivariational}.
The mean-field energy is composed of five terms:
\begin{align}
E_{\rm HFB} &= E_{\rm kin} + E_{\rm Sk} + E_{\rm pair} + E_{\rm Coul}  + E^{(1)}_{\rm cm}\, ,
\label{eq:Ehfb}
\end{align}
which are, respectively, the contributions of the kinetic energy,
 the Skyrme effective interaction
\cite{Bartel82,Sharma95,Reinhard95,Kortelainen10,Chabanat98,Kortelainen12}, 
a zero-range pairing interaction with appropriate cutoffs~\cite{Krieger90}, 
the Coulomb force~\cite{Brown00,Goriely08}, and the one-body part of 
the centre-of-mass correction. 
The correction energy $E_{\rm corr}$ consists of three parts:
\begin{align}
E_{\rm corr}&= E_{\rm rot} + E^{(2)}_{\rm cm} + E_{\rm W}  \, ,
\label{eq:Ecorr}
\end{align}
which are, respectively, the rotational correction
~\cite{Belyaev61,Goriely07,Pena16}, the two-body part of the centre-of-mass correction
~\cite{Bender99}, and the Wigner energy \cite{Goriely13b}. 
In what follows, we will describe each term in more detail separately, but
we emphasize here that all of the ingredients of these terms are calculated 
consistently from the auxiliary state $|\Phi\rangle$.

\subsubsection{The Skyrme energy}

The total Skyrme energy can be written in terms of an integral over 
an energy density $\mathcal{E}_t(\bold{r})$, where $t=0,1$ is the isospin index,
as
\begin{align}
\label{eq:Eskyrme}
E_{\rm Sk} &= \int d^3 \bold{r} \sum_{t = 0,1} \mathcal{E}_t(\bold{r}) \, .
\end{align}
The energy density is given by
\begin{align}
\label{eq:Skyrme}
\mathcal{E}_t(\bold{r})
& =  
              C^{\rho\rho}_t \, \rho_t^2 (\bold{r})
            + C^{\rho\rho\rho^{\gamma}}_t \rho_0^\gamma (\bold{r}) \, \rho_t^2 (\bold{r}) \nonumber \\
&           + C^{\rho\tau}_t \, \rho_t (\bold{r}) \, \tau_t (\bold{r})
            + C^{\rho \Delta \rho}_t \, \rho_t (\bold{r}) \, \Delta \rho_t (\bold{r})  \nonumber \\
&           + C^{\rho \nabla \cdot J}_t  \rho_t (\bold{r}) \, \boldsymbol{\nabla} \cdot \bold{J}_t (\bold{r}) \, , 
\end{align}
in terms of three local, time-even 
densities $\rho_t(\bold{r})$, $\tau_t(\bold{r})$ and 
$\bold{J}_t(\bold{r})$ that characterize the auxiliary state $|\Phi\rangle$~\cite{Ryssens15}. 
The ten coupling constants
\{C\} are 
determined in terms of
the model parameters $t_{0-3}$, $x_{0-3},$ $W_0,$ and $W_0'$ (see 
\ref{app:couplingconstants} for more details).

The functional form of Eq.~\eqref{eq:Skyrme} is closely related (but not 
equivalent) to the form generated by a (density-dependent) Skyrme effective 
interaction, and is fairly standard in that respect. Nevertheless, we consider 
three aspects of Eq.~\eqref{eq:Skyrme} worthy of discussion. First, we employ 
an extended spin-orbit term as originally introduced in 
Refs.~\cite{Sharma95,Reinhard95} resulting in the presence of two spin-orbit 
parameters $W_0$ and $W_0'$ instead of just one with
$W_0 = W_0'$ as was done for the majority of Skyrme EDF parameterizations.
Other examples using this extended form are the series of UNEDF interactions 
\cite{Kortelainen14,Kortelainen10,Kortelainen12} and those of Ref.~\cite{Klupfel_09}.
Second, as with the BSk forces after BSk19 \cite{Goriely16,Goriely13a,Goriely13b},
SLy4~\cite{Chabanat98}, SkM*~\cite{Bartel82}, UNEDF0~\cite{Kortelainen10} and
UNEDF1~\cite{Kortelainen12},
Eq.~\eqref{eq:Skyrme} does not include the terms bilinear in the spin-current 
tensor density $J_{\mu\nu}(\bold{r})$ that are obtained when evaluating the 
expectation value of an effective two-body Skyrme interaction~\cite{Ryssens15}.
Third, we do not consider any terms involving time-odd densities for finite 
nuclei, as all such contributions vanish identically for the configurations
we consider during the model adjustment, see Sec.~\ref{sec:Symmetries}.

\subsubsection{The pairing energy} To incorporate the effect of pairing correlations, 
we include a simple pairing term of the form
\begin{align}
E_{\rm pair} &= \sum_{q=p,n} \frac{V_{\rm \pi q}}{4} \int d^3\bold{r}
\, \left[1 - \eta \left(\frac{\rho_0(\bold{r})}{\rho_{\rm sat}}\right)^{\alpha} \right]  
\tilde{\rho}_q^*(\bold{r}) \tilde{\rho}_q(\bold{r})
\, ,
\end{align}
where $\rho_{\rm sat} = 0.16$ fm$^{-3}$ and $\tilde{\rho}_q(\bold{r})$ is the 
local pairing density~\cite{Dobaczewski84}.
The parameter $\eta$ changes the character of the pairing 
interaction from being mainly active in the nuclear volume ($\eta = 0$) to being
limited to the nuclear surface region ($\eta = 1$).

The contribution of each individual single-particle state to the pairing 
density $\tilde{\rho}_q(\bold{r})$ 
is weighted with a cutoff factor $f^{\rm pair}_{q,i}$, limiting the 
pairing interaction to levels in an energy window around the Fermi energy 
$\lambda_q$. These factors take the form~\cite{Krieger90}
\begin{align}
f^{\rm pair}_{q,i} &= \prod_{\sigma = \pm 1}
     \left[ 1 + e^{ \sigma \left( \epsilon_i - \lambda_q 
                             - \sigma E_{\rm cut}\right) /\mu_{\rm pair}}  \right]^{-1/4} \, ,
\label{eq:paircut}
\end{align}
where $\epsilon_i$ is the single-particle energy of the $i$-th single-particle
state in the basis that diagonalizes the single-particle Hamiltonian,
$\mu_{\rm pair} = 0.5$ MeV and $E_{\rm cut}$ is the energy cut-off, which is an 
adjustable parameter of the model.

\subsubsection{The Coulomb energy}
\label{sec:Coulomb}
We only take into account the direct contribution of the Coulomb interaction to 
the energy
\begin{align}
\label{eq:Ecoul}
E_{\rm Coul} &\equiv E^{\rm direct}_{\rm Coul} = 
                        \int d^3\bold{r} \, U(\bold{r}) \rho_c(\bold{r}) \, , 
\end{align}
where $\rho_c(\bold{r})$ is the charge density of the nucleus. $U(\bold{r})$
is the Coulomb potential satisfying the electrostatic Poisson equation for the 
charge density $\rho_c(\bold{r})$
\begin{align}
\Delta U(\bold{r}) &= - 4 \pi e^2 \rho_c(\bold{r}) \, ,
\label{eq:poisson}
\end{align}
with $e^2$ the square of the elementary charge.  To account for the finite size
of the charge distributions of individual protons and neutrons, the charge 
density in Eqs.~\eqref{eq:Ecoul} and \eqref{eq:poisson} is constructed through 
the folding of (point) proton and neutron densities with appropriate form 
factors. For protons, we employ the Gaussian form factor of  
Ref.~\cite{Negele70}, characterized by the rms radius of the proton 
$r_p = 0.895$~fm of Ref.~\cite{Sick03}. For neutrons, we employ the difference 
of two Gaussians of widths $r_+^2=0.387$ fm$^2$ and 
$r_{-}^2=0.467$ fm$^2$~\cite{Chandra76}, corresponding to a neutron mean 
square charge radius of $-0.116$ fm$^2$. 

As a practical recipe to simulate beyond mean-field Coulomb correlations, we 
drop the exchange contribution of the electrostatic interaction entirely. This 
recipe was first studied in Ref.~\cite{Brown00}, and employed in the HFB-series 
of mass models, starting with HFB-15~\cite{Goriely08}. 

\subsubsection{The Wigner energy}
The Wigner energy is a phenomenological term included 
to simulate the excess binding energy of $N \sim Z$ nuclei. We take the 
form introduced in Ref.~\cite{Goriely13b}:
\begin{align}
E_{\rm W}=& V_W\exp(-\lambda((N-Z)/A)^2) \nonumber \\
         &+ V_W^{\prime}|N-Z|\exp\left[-(A/A_0)^2\right] \, ,
\label{eq:wig}
\end{align}
which depends on four parameters $V_W, V_W', \lambda$, and $A_0$.
This correction chiefly influences light and $N\simeq Z$ nuclei.

\subsubsection{Centre-of-mass correction}
Although the details vary, the inclusion of some correction for spurious motion 
of the centre-of-mass of the nucleus is standard practice in the 
literature~\cite{Ben03}. This correction reads

\begin{align}
E_{\rm cm} &=      - \frac{1}{2mA} \langle \hat{\bold{P}}^2_{\rm cm}\rangle 
            \equiv E^{(1)}_{\rm cm} + E^{(2)}_{\rm cm} \, ,
\end{align}

\noindent where $m$ is the (average) nucleon mass and $A$ is the mass number. We split 
this correction into two separate contributions. 
The one-body part can be taken into account as a rescaling of the 
kinetic energy~\cite{Ben03}, but the calculation of the two-body 
contribution is much more complicated, as described in Ref.~\cite{Bender99}.

\subsubsection{The rotational correction}
Less standard in the literature is the inclusion of a rotational correction, 
whose goal is to simulate the effect of the restoration of rotational symmetry.
Inspired by the HFB mass models of Ref.~\cite{Goriely16} and references therein,
we employ a correction based on a simple perturbative cranking model, involving the 
Belyaev moments of inertia (MOI) around 
the three principal axes of the nucleus,
$\mathcal{I}_{\mu}$($\mu=x,y,z$)~\cite{Belyaev61,Ryssens15}:

\begin{subequations}
\begin{align}
\label{eq:Erot}
E_{\rm rot} &= - 
\sum_{\mu=x,y,z} f_{\mu}^{\rm rot}\frac{\langle \hat{J}^{2}_{\mu} \rangle}{2 \mathcal{I}_{\mu}} \, , \\
f_{\mu}^{\rm rot} &=  b  \tanh \left( c \frac{\mathcal{I}_{\mu}}{\mathcal{I}_c}\right) \, ,
\label{eq:rotcut}
\end{align}
\end{subequations}
where $\hat{J}_{\mu}$ is an angular momentum operator and 
$\mathcal{I}_{\rm C} = \tfrac{2}{15} m R^2 A$ is (one third of) the 
MOI of a rigid rotor of radius $R = 1.2 A^{1/3}$, comprised of $A$ 
nucleons of average mass $m = \tfrac{1}{2} (m_n + m_p)$. The three
MOI in Eq.~\eqref{eq:Erot} are obtained consistently from the 
auxiliary state $|\Phi\rangle$, but their calculation is not trivial, and we
refer to \ref{app:rotational} for a more detailed discussion.

The inclusion of a cutoff factor $f_{\mu}^{\rm rot}$ in Eq.~\eqref{eq:Erot}
is necessary to smooth the transition between deformed and spherical nuclei. 
In early tests, we employed the smoothing prescription of 
Refs.~\cite{Goriely07,Pena16}, but found it unsuitable for triaxial 
systems. Although they are of similar size, our parameters $b$ and $c$ are therefore
not directly comparable to their counterparts in
Refs.~\cite{Goriely07,Pena16}.


\subsection{Self-consistent HFB calculations in coordinate-space}
\label{Sec:HFB_calculations} 

\subsubsection{Coordinate-space representation: the MOCCa code}

The variation of the mean-field energy $E_{\rm HFB}$ leads to the 
  self-consistent Skyrme-HFB equations, which need to be solved iteratively~\cite{Ryssens15,RingSchuck}.
  To this end, we employ the MOCCa code~\cite{Ryssens16,MOCCa,Rys19}, 
which represents the single-particle wave functions on a three-dimensional 
Cartesian Lagrange mesh~\cite{Baye86,Gall93,Bonche05,Baye15,Ryssens15}. This 
representation allows us to treat different nuclear shapes on an equal footing 
and yields a high numerical accuracy that is essentially independent of 
deformation~\cite{Ryssens15b}.  
 
The MOCCa code is similar in spirit to the earlier EV8 
code~\cite{Bonche05,Ryssens15}, but differs from it in four respects. First, 
MOCCa allows the user significant freedom of choice with respect to 
self-consistent symmetries and is capable of completely symmetry-unrestricted 
calculations. Second, it further improves on the numerical accuracy of EV8 
through the self-consistent use of Lagrange derivatives instead of employing 
finite difference formulas on the mesh~\cite{Ryssens15b}. Third, MOCCa's further algorithmic improvements have made its iterative process faster than  EV8's by up to an order of magnitude~\cite{Rys19}. Finally, 
MOCCa is equipped to handle pairing correlations at the HFB level, employing 
the two-basis method~\cite{Gall93,Rys19}.

Throughout the adjustment process, we employed cubic meshes with
$N_x=N_y=N_z=32$ points and a mesh spacing $dx = 0.8$ fm. During the final calculation, we expanded the lattice to $N_x=N_y=N_z=36$. With these numerical 
choices, we can estimate that the calculated binding energies and rms radii will
 not change by more than a few tens of keV and 0.01 fm respectively for the 
 majority of nuclei when further expanding the numerical 
basis~\cite{Ryssens15b}. Achieving comparable accuracy using basis-expansion
methods requires the inclusion of a number of harmonic oscillator states that is 
prohibitive for calculations on the scale of the nuclear chart, for both Skyrme
~\cite{Carlsson10} and Gogny EDFs~\cite{Arzhanov16}.

To avoid the extreme memory requirements of representing a complete set of single-particle states on 
coordinate meshes of this size, we store and iterate only the $N_N$ neutron
states and $N_Z$ proton states with lowest single-particle energy. On the order
of a few hundred single-particle states suffices to calculate all relevant 
quantities, which are all computed from the single-particle states weighted by 
their occupation numbers\footnote{With the exception of the Belyaev MOI, 
see \ref{app:rotational}.}. For a given nucleus, we iterated $N_N = N + 160$ neutron and $N_Z = Z + 100$ proton states during the 
adjustment process. For the final calculation of the mass table we increased these numbers to $N_N = N + 400$ neutron and $N_Z = Z + 240$ proton 
states.

Previous BSk models were adjusted using an implementation that relied on an 
  expansion of the single-particle wave functions in a set of harmonic 
  oscillator basis states. An important difference between this type
  of numerical representation and a coordinate space one is the treatment of
  pairing. First, the discretization of the continuum of both approaches is
  vastly different, resulting in a different spectrum of positive-energy 
  single-particle states. Second, the limited number of single-particle states
  we iterate forces us to limit the width of the pairing window, resulting in 
  values of $E_{\rm cut}$ in Eq.~\eqref{eq:paircut} that are significantly
  smaller than those typically used in harmonic oscillator approaches.
 The combination of these differences makes it essentially impossible to 
 directly use the BSk functionals in our coordinate space representation, and 
 clearly indicates the need for a new fit.

\subsubsection{Imposed symmetries: nuclear configurations considered}
\label{sec:Symmetries}

EDF-based nuclear models rely on the notion of intrinsic symmetry
    breaking to achieve their descriptive power. It is in principle desirable 
    to perform completely symmetry-unrestricted calculations, {\it i.e.}\ to consider 
    the most general nuclear configurations without any restrictions.
Although the MOCCa code can perform such symmetry-unrestricted 
calculations, those come at a price: (i) they significantly increase the computational cost and (ii) they imply the loss of all quantum
numbers, making interpretation of results and comparison to experiments difficult. 
Techniques to restore broken symmetries and recover the associated quantum 
numbers for triaxial configurations exist \cite{Bally2021}, but are beyond the scope of this 
contribution. 

In fact, there is a large body of empirical evidence that the HFB ground states of
all nuclei usually adopt one or several spatial symmetries, which can be used to simplify 
the numerical treatment by imposing those that remain conserved for most, if not all, nuclei. 
We have restricted ourselves here to nuclear configurations that respect three 
plane-reflection symmetries, as well as time-reversal symmetry. This choice 
of spatial symmetries results in nuclear shapes that are reflection-symmetric and
are invariant under discrete rotations of $180^{\circ}$ around any principal axis.
If visualized, virtually
all configurations discussed here would resemble ellipsoids with three principal 
axes of (possibly) different lengths. This resemblance is not exact, 
and large numbers of nuclei exploit non-zero values of higher-order multipole 
deformations beyond quadrupole, as will be discussed below. 

A nucleus with non-axial deformation does not exhibit any continuous 
rotational symmetry, and we consequently cannot assign any definite 
rotational quantum number. The only non-trivial quantum number we can assign 
without ambiguity for all configurations considered is parity. Eliminating 
this restriction would allow us to study reflection asymmetric nuclear 
configurations, which are typically characterized by a non-zero octupole 
deformation. Such configurations are of importance for the description of 
fission barriers~\cite{Goriely07,Bender20}, but systematic calculations 
have shown that static octupole deformation is only expected for a limited 
number of nuclides~\cite{Robledo2011,Agbe16,Cao_2020}. While the effect on the 
binding energy for these isotopes can be sizeable (up to 2 MeV), we have opted 
to not explore octupole deformation for this study. Imposing time-reversal 
symmetry on the other hand chiefly limits our description of odd-$A$ and odd-odd nuclei, which we will discuss in more detail in the next section. But it 
can already be noted here, however, that the contribution from these terms to the 
total binding energy remains on the order of at most a few hundred 
keV~\cite{Schunck10,Pototzky10} and thereby remains smaller than the average
deviation of nuclear masses that we achieve in the parameter adjustment. 

Lifting the restrictions imposed by reflection symmetry and time-reversal 
invariance is feasible, and the necessary preparations for doing so in future mass fits are underway. 
We note that the expected increase in complexity due to these generalizations
is significantly smaller than the one incurred when generalizing from two-dimensional
axially-symmetric shapes to three-di\-men\-sional triaxial shapes, as we explore
here.

The shape of the nuclear density can be characterized in terms of 
multipole moments $\hat{Q}_{\ell m} \equiv \hat{r}^{\ell} \hat{Y}_{\ell m}$, 
where $\hat{Y}_{\ell m}$ is a spherical harmonic. With the symmetries chosen here, 
all multipole moments $\langle\hat{Q}_{\ell m} \rangle$ are real, and can 
take finite values only when $\ell$ and $m$ are both even. The multipole 
moments are most transparently discussed in terms of dimensionless 
deformation parameters $\beta_{\ell m}$, which we define as
\begin{align}
\beta_{\ell m} &= \Re \left( \frac{4 \pi } {  3  (r_0 A^{1/3})^{\ell}  A }  \langle \hat{Q}_{\ell m} \rangle \right) \, ,
\label{eq:betalm}
\end{align}
where $r_0=1.2$ fm. We emphasize that these multipole moments characterize the 
shape of the nuclear \emph{volume}, rather than the deformation of the 
nuclear \emph{surface}. The latter type of moments is generally employed in 
microscopic-macroscopic approaches, such as the $\epsilon_{\ell}$ defined in 
Ref~\cite{Moller06}. 

For states obtained from a self-consistent minimization it 
is not unusual to find numerically significant multipolarities as large as 
$\ell = 10$, but we will mainly discuss the quadrupole ($\ell = 2$)
moments, $\langle\hat{Q}_{20} \rangle$ and $\langle \hat{Q}_{22}\rangle$
as they
represent the dominant deformation modes. 
The nuclear quadrupole deformation is also often discussed
in terms of the total size of the deformation $\beta$ and the triaxiality
angle $\gamma$:

\begin{subequations}
\begin{align}
\label{eq:beta}
\beta  &= \sqrt{\beta_{20}^2 + 2 \beta_{22}^2 } \, , \\
\gamma &= \text{atan} \left( \sqrt{2}\beta_{22}/ \beta_{20} \right) \, . 
\label{eq:gamma}
\end{align}
\end{subequations}
With the symmetries imposed on our calculations, we can limit the discussion to one sextant of the $\beta$-$\gamma$ plane. For finite values of $\beta$, prolate shapes correspond to $\gamma = 0^{\circ}$
while oblate shapes correspond to $\gamma = 60^{\circ}$. The introduction of the
triaxial degree of freedom allows the nucleus to explore all values of $\gamma$
in between these two extremes.

 The size of the quadrupole deformation, $\beta$, is a rotational invariant. 
    For higher order multipole deformations, we can similarly define rotational
    invariants $\beta_{\ell}$ as 
\begin{align}
\beta_{\ell} &= \sqrt{ \sum_{m=-\ell}^{\ell} \beta^2_{\ell m} }\, .
\label{eq:sigmadef}
\end{align}
The value of $\beta_{\ell}$ does not specify completely the deformation of 
   the nucleus at order $\ell$, and only for $\ell = 2$ does the value of the single additional quantity ($\gamma$) suffice to do so. 

From a practical point of view, the three conserved spatial symmetries allow us to limit 
the calculations to a mesh of effective dimensions $(N_x/2, N_y/2, N_z/2)$ 
while we can exploit time-reversal to limit the effective calculation to 
$N_N/2$ and $N_Z/2$ single-particle states~\cite{Ryssens15}, reducing 
the computational burden in both CPU time and memory required by a factor of
eight compared to the most general possible calculation.  

\subsubsection{Nuclei with odd nucleon number(s)}
\label{sec:oddnuclei}

The fully self-consistent treatment of odd-$A$ and odd-odd nuclei in the context of HFB 
theory requires the construction of one or two quasiparticle excitations with 
respect to a reference state of even-even character. Because of polarization effects,
this blocking procedure results in an auxiliary HFB state that is no longer 
invariant under time-reversal, which in turn implies the need to consider terms
involving time-odd densities in the Skyrme EDF~\cite{Ben03}. 

As we focus in this study on the impact of non-axial shapes and using a 
  coordinate-space representation, we opted  to side-step this complexity,
employing the equal-filling method~\cite{Perez08} to construct statistical mixtures of
Bogoliubov reference states that are manifestly time-reversal invariant.
This approximation takes into account the blocking effect due to 
the odd nucleon(s), but neglects polarization effects due to time-odd terms of 
the EDF.

The blocking of quasiparticles can render self-consis\-tent HFB 
calculations notoriously difficult to converge~\cite{Schunck10}. The chief 
reason for this is the need to select, at every self-consistent iteration, the 
appropriate quasiparticle(s) to block. This procedure is to a certain degree 
robust, if one targets states that are characterized by a set of quantum
numbers or, more generally, by fixed expectation values of one or more operators.
In our case, 
however, we are interested in the overall lowest binding energy 
\emph{after self-consistency is achieved}, independent of the characteristics
of the auxiliary HFB state. 

Our strategy consists of blocking the quasiparticle with the lowest
quasiparticle energy at every iteration. Nevertheless,
from any given iteration to the next, the candidate quasiparticle excitation 
can change dramatically in character. To limit somewhat the destructive
influence of crossings in the quasiparticle energies, we employ parity as the 
sole remaining quantum number. For an odd-$A$ nucleus we thus perform two 
calculations, constructing both the lowest blocked states of positive and 
negative parity. For an odd-odd nucleus, we perform four calculations to 
construct all possible combinations of one-proton-one-neutron excitations 
for a given parity of each species that are lowest in energy. 
The parity corresponding to the lowest overall energy is taken as that of the 
ground state. While computationally costly, this strategy results in at least 
one converged calculation for virtually all nuclei throughout the adjustment.

Finally, a comment on the angular momentum quantum number $J$ is in order. The quantum
    numbers $J^{\pi}$ of the ground states of odd-$A$ and odd-odd nuclei are of great 
   interest, but, as mentioned above, we can only assign the parity quantum number $\pi$. 
   We have made no effort to 
   extract any predictions for $J$ from our calculations due to the lack of 
   rotational quantum numbers. In principle, one could employ a simple model 
   to extract $J$ from a symmetry-broken mean-field calculation, such as a 
   strong-coupling model in the case of an axially deformed configuration. 
   However, this 
   type of argument is generally only applicable in limited regions of the 
   nuclear chart (usually heavy, well-deformed and axially symmetric nuclei).
   We are not aware of any recipe that is globally applicable, except for the extremely
   demanding symmetry restoration techniques that are beyond the scope of the present global study.

\subsubsection{Minimization of the energy}
\label{sec:semivariational}

Ideally, one would like to employ the variational principle and minimize the 
   total energy (Eq.~\ref{eq:Etot}).
   The formal variation of $E_{\rm HFB}$ is rather straightforward, and 
   numerical minimizations of this quantity are performed routinely nowadays. 
   The self-consistent equations become significantly more involved if 
   the two-body centre-of-mass and rotational corrections are included in the
   optimization. The variation of $E_{\rm cm}^{(2)}$ 
   has only rarely been performed consistently, exceptions being
   the SLy6 and SLy7 parameterizations of Ref.~\cite{Chabanat98}.
   To the best of our knowledge, the consistent variation of the rotational 
   correction $E_{\rm rot}$ has never been attempted.

  Traditionally, if a two-body centre-of-mass correction or a rotational 
  correction are included in the model, they are treated \emph{perturbatively}:
  one calculates the total energy $E_{\rm tot}(|\Phi_0\rangle)$ from the 
  auxiliary state $|\Phi_0\rangle$ which minimizes the mean-field energy
  $E_{\rm HFB}$ only. While easy to implement, a perturbative approach 
  suffers greatly if two (or more) coexisting auxiliary states
  have a quasi-identical mean-field energy $E_{\rm HFB}$ but different 
  $E_{\rm corr}$. In such cases, a perturbative approach will result in
  disproportionally large changes in the total energy $E_{\rm tot}$ when 
  either varying the nucleon number or (slightly) varying the parameters of 
  the Skyrme interaction, leading respectively to unphysical separation energies 
  or convergence problems for the fitting procedure.
  
Instead of a perturbative treatment, we have employed a 
  \emph{semivariational} strategy to include these corrections. For a given 
  nucleus, we perform a large number of calculations that include 
  $E^{(2)}_{\rm cm}$ and $E_{\rm rot}$ perturbatively, each constrained
  to different values of the quadrupole deformation.
  Our final value for the binding energy is then selected as the overall 
  minimum of the total energy $E_{\rm tot}$ as a function of quadru\-pole 
  deformations with a resolution of $\Delta \beta_{20} =\Delta \beta_{22} =0.005$. 
  We emphasize that we also employ this strategy for blocked calculations:
   we scan the full $\beta$-$\gamma$-plane for each type of quasiparticle 
   excitations considered, resulting in two scans for odd-$A$ and four scans
   for odd-odd nuclei. 

Despite the evident computational complexity of this approach,
   we have adopted the semivariational strategy for three 
   reasons. First and foremost, early tests showed a systematically improved 
   description of the binding energies. Second, the 
   inclusion of quadrupole constraints increases the stability of blocked
   calculations by limiting the amount of possible 
   single-particle level crossings in any single calculation (see also the previous section).
   Finally, this approach greatly alleviates the problem associated with 
   coexisting mean-field minima discussed above.

\section{Model adjustment using neural networks}
\label{Sec:Fit} 
Our mass model depends on 22 parameters: eleven are related to the Skyrme
EDF ($t_0, t_1, t_2, t_3, x_0, x_1, x_2, x_3,$ $W_0, W_0', \gamma$),
five  to the zero-range pairing interaction 
($V_{\pi n}, V_{\pi p}, E_{\rm cut}, \eta, \alpha)$,
 two to the rotational correction ($b, c$),
and four to the Wigner energy ($V_W, \lambda, V_{W}', A_0$). 
During the fitting procedure, we set $\gamma=0.3$  to ensure the correct 
description of the nuclear matter incompressibility and impose a symmetry energy
at saturation density of $J=32$~MeV to ensure a minimal stiffness of the neutron matter 
equation of state \cite{Goriely16}. 

Referring to the parameters as $\boldsymbol{\chi} = (t_0, t_1, \ldots)$, our 
goal is to minimize an objective function $\sigma_{\rm rms}(\boldsymbol{\chi})$,
defined for several observables $O$ as 
\begin{align}
\sigma_{\rm rms}(\boldsymbol{\chi}) &= \sqrt{\sum_{N,Z,O}\left[ W_{O} \; \sigma(N,Z,\boldsymbol{\chi},O) \right]^2 } \, ,
\label{eq:sigmadef}
\end{align}
in terms of weights $W_O$ for each observable and the deviations between 
experimental and calculated values 
$\sigma(N,Z,\boldsymbol{\chi}, O ) =  O_{\rm exp}(N,Z) - O_{\rm th}(N,Z,\boldsymbol{\chi})$.
The objective function considered here includes all the 2408  binding energies 
known experimentally for $Z,N \ge 8$ nuclei \cite{Wan17}, making even a single 
evaluation of Eq.~\eqref{eq:sigmadef} a demanding task. 

In principle, we could proceed to minimize the objective function using 
traditional minimization methods. However, any systematic search of the 
high-dimen\-sional parameter space is prohibitively expensive.
The problem is further complicated by the nonlinearity of the self-consistent 
Skyrme-HFB equations: it is not easy to predict the results of even small 
variations of individual parameters.

In order to explore the parameter space efficiently, we have developed a new
   approach inspired by Ref.~\cite{Las20}. We consider a committee of Multi-Layer
   Neural Networks (MLNNs)~\cite{Cho17} as an emulator, {\it i.e.}\ as a computationally 
   cheap estimate of the observables as a function of $N,Z$ and $\boldsymbol{\chi}$. 
Each individual MLNN aims to provide an estimate
for the output of the code, \textit{i.e.}\ an estimate 
$\sigma'(N,Z,\boldsymbol{\chi}, O )$ of the actual $\sigma(N,Z,\boldsymbol{\chi}, O)$. 
The minimum of the estimated objective function $\sigma'(\boldsymbol{\chi})$
can be found at little computational cost, resulting in a prediction for an optimal
parameter set by each member of the committee. Until convergence of the value
of the objective function, the MLNNs are trained on an increasing set of MOCCa 
calculations in regions of the parameter space deemed promising by the committee. 

During the optimization procedure, our committee typically consists of
$N_{\rm NN} = 300$ neural networks. Each individual member consists of four
hidden layers of neuron number\footnote{All the neurons have a rectifier activation function $f(x)=\max(0,x)$.} 128, 64, 32, and 16. 
The output layer aims at estimating the difference between theoretical and 
experimental values for a given observable, based on the values of $N$, $Z$ 
and $\boldsymbol{\chi}$. To aid the learning process, we provide each member 
with additional information: the mass number $A$, the number parity of both 
$N$ and $Z$, the distance to the closest magic number (up to 126) and individual
terms of the liquid drop model for the nuclear binding energy 
($A^{2/3}$, $Z(Z-1)/A^{1/3}$, $(N-Z)^2/A$, and $A^{-1/2}$).

Every MLNN is initialized randomly and trained on available data obtained from 
MOCCa calculations (see below) using the Keras/Tensorflow libraries~\cite{Cho17}. 
Once trained, we minimize the objective function (as modelled by each member of 
the committee) with respect to the mass model parameters using the trust region 
reflective algorithm \cite{Byr88}. This results in a predicted optimal
parameter set $\boldsymbol{\chi}^i$, one for each member of the committee
($i=1, \ldots, N_{\rm NN}$). As each member is initialized differently, 
they will provide slightly different predictions and so propose different 
candidate parameter sets.
  
The MLNNs are trained on an always-increasing library of MOCCa calculations,
{\it i.e.}\ an individual data point is the difference between the experimental value 
of an observable and the calculated value for a given nucleus and a value of 
$\boldsymbol{\chi}$. To start the learning procedure, we compute a first set of 
$N_{\rm ini} \sim 1000$ such data points, for essentially random 
 nuclei within the AME2016 database with $Z\geq8$, and $N\geq8$  and values of the mass model parameters in 
a given range. When possible these limits are taken from constraints 
imposed on nuclear matter properties~\cite{Goriely13b,Goriely16} and obtained
by reasonable guess otherwise. 

After the initial training, we keep growing the training library
    guided by the predictions of individual commitee members. For each 
    candidate parameter set $\boldsymbol{\chi}^i$, we calculate observables for 
    three random nuclei, resulting in $3N_{\rm NN}$  extra data points. 
    Continuing the data set
    training on the expanded data set, we generate new candidate parameter sets. 
    This process is continued until convergence, {\it i.e.}\ until 
    we achieve no significant further decrease of the objective function.  

 Periodically, we interrupt this individual phase of the optimization 
    strategy to include either (i) a collective decision of the committee 
    or (ii) an active learning step.  Every time the data set is increased by 
    $N_{\rm coll} \sim 8000$ data points, we perform a collective step. 
    When the data set is increased by a further $N_{\rm act} \sim 4000$ 
    points, we perform an active learning step.

 A collective step is started by polling
    every member of the committee on the candidate parameter sets of 
    \emph{the other members}. Every candidate parameter set
    is then assigned a collective estimated deviation 
    $\sigma^{\rm coll}_{\rm rms}(\boldsymbol{\chi}^{i})$, computed as the 9th
    decile of the modelled deviations for this parameter set among all members 
    of the committee.\footnote{One could imagine more straightforward recipes to assign a 
    collective score, such as the average of all modelled deviations, but we
    have found these to be not very reliable in early tests.} 
    Among the $N_{\rm NN}$ candidate parameter sets, we select
    the one with the lowest collectively-estimated deviation and use it to 
    perform MOCCa calculations for all nuclei ({\it i.e.}\ 2408 data points). Once 
    these results are added to the data set, the individual training phase 
    resumes.

 In an active learning step, we identify 100 pairs of nuclei and parameters sets whose predicted 
    contribution to the objective function is the largest for every candidate
    parameter set $\boldsymbol{\chi^i}$. We perform MOCCa calculations for this 
    set of nuclei-interaction pairs, 
    thereby generating data that is likely to improve the overall predictions
    of the committee members. 
 
 We note that the Wigner energy $E_{\rm W}$ was not included
   in the machine-learning protocol. Since it is a simple analytical function
   of $N,Z$ and the four parameters $(V_W, V_W', \lambda,A_0)$, we have simply
   added it manually to the output of each MLNN. This reduces the difficulty of 
   the learning, as the committee members do not need to know the dependence of 
   the results on these parameters. 

After convergence we performed a full calculation for all nuclei with 
   $8 \leq Z \leq 110$ from the proton- to the neutron-drip line using the final parameter set.
    All the results presented below are generated with the MOCCa code, none of them are predictions by a neural network.

\section{The BSkG1 parameterization}
\label{Sec:BSKG1} 

\begin{table}[t]
\caption{BSkG1 parameter set:
         16 parameters characterizing the self-consistent mean-field energy $E_{\rm HFB}$
         and 6 corresponding to the correction energy $E_{\rm corr}$.
}
\begin{tabular}{l|d{8.6}}
\hline
\hline
 \mc{Parameters}  &  \mc{ BSkG1 }  \\
\hline
$t_0$ [MeV fm$^3$]  & -1882.36    \\
$t_1$ [MeV fm$^5$] &  344.79   \\
$t_2$ [MeV fm$^{5}$] & -2.43198   \\
$t_3$ [MeV fm$^{3 + 3\gamma}$]  &  12322.0   \\
$x_0$ &  0.196276 \\
$x_1$ &  -0.580308   \\
$x_2 t_2$ [MeV fm$^5$]  & -170.203  \\
$x_3$ &  0.120751  \\
$W_0$  [MeV fm$^5$]&   123.922   \\
$W_0'$ [MeV fm$^5$] & 83.519  \\
$\gamma$  & 0.3  \\
\hline
$V_{\pi n}$ [MeV] & -644.921  \\
$V_{\pi p}$ [MeV] & -682.559 \\
$\eta$ &  0.692  \\
$\alpha$ &  0.77 \\
$E_{\rm cut}$ [MeV]&  7.42  \\
\hline
\hline
$b$ &  0.93 \\
$c$ & 5.00  \\
\hline
$V_W$ [MeV] &  -1.905  \\
$\lambda$ &  272.2  \\
$V_W'$ [MeV] &  0.671   \\
$A_0$ &  36.211 \\
\hline
\hline
\end{tabular}
\label{tab:param_skyrm}
\end{table}

\subsection{Ingredients of the objective function and parameter values}
\label{sec:ingredients}

As mentioned above, the objective function 
(Eq.~\ref{eq:sigmadef}) includes all the 2408 measured masses for $Z$, $N\ge8$ nuclei compiled with the 
2016 atomic mass evaluation (AME2016) database~\cite{Wan17}.
However, it is well known that fits based solely on nuclear masses 
generally lead to an excessive pairing strength~\cite{Goriely06}, hence
to unreliable extrapolations towards unknown regions of the nuclear 
chart. To avoid such a shortcoming, it is important to either constrain the 
pairing strength or include additional observables in the objective function. 
To that aim, we added to the objective function the MOI of
heavy nuclei deduced from observed rotational bands. This set
consists of 48 even-even nuclei, most of which are neutron-rich rare-earth 
nuclei~\cite{Zeng94,Afanasjev2000,Pearson1991}. We compare the experimental
values with the calculated Belyaev MOI (see also \ref{app:rotational}), 
which are highly sensitive to the strength of the pairing interaction.

Following the Brussels-Montr\'eal protocol, charge radii as well as nuclear 
matter properties are also qualitatively included in the fitting strategy, 
though they are not explicitly included in the objective function.
This was achieved by constraining some of the nuclear matter properties, as followed in Ref.~\cite{Goriely16}, namely
(i) the symmetry coefficient is set to $J=32$ MeV to ensure a certain degree
of stiffness of the infinite neutron-matter equation of state \cite{Goriely16}; (ii) the Fermi 
wave number $k_F$ is determined to best reproduce nuclear charge radii; (iii) the exponent $\gamma$ is set to 0.3 
to ensure the incompressibility 
$K_{v}$ of charge-symmetric infinite nuclear matter lies within the interval 
$230 \leqslant K_{v} \leqslant 250 $~MeV \cite{Colo04}; and (iv) the isoscalar effective mass $M_s^*/M$ at the saturation density $n_0$
is taken close to the value of 0.84, as predicted by extended Brueckner-Hartree-Fock (EBHF) calculations~\cite{Cao06,Zuo02}.

Our committee-guided adjustment procedure proved particularly practical
   in exploring the compromise between the reproduction of all 
   nuclear masses and the reproduction of charge radii, realistic pairing, and nuclear matter properties. Once the members are
   sufficiently trained, the committee can propose different parameterizations 
   (and estimate their properties)
   as a function of the weights $W_O$ at low computational cost. For instance,
   this allowed us to thoroughly explore the trade-off between masses and MOI
   by varying the relative weights $W_{\rm MOI}/W_{\rm mass}$ without the 
   need to restart the fit multiple times. 

We refer to the final functional of the adjustment procedure as BSkG1, 
   the parameters of which are shown in Table \ref{tab:param_skyrm}. The corresponding infinite nuclear matter properties are given in Table \ref{tab:prop}.    
   The rms $\sigma$ and mean deviation $\bar{\epsilon}=  \overline{ O_{\rm exp}(N,Z) - O_{\rm th}(N,Z)}$ 
   for both the masses and charge radii are given in Table \ref{tab:rms}. Note that the model was
   adjusted to the AME2016 masses~\cite{Wan17} leading to an rms deviation of 0.734~MeV with respect to all the
   2408 masses. However, the deviations given in Table \ref{tab:rms} 
   have been calculated with respect to the (slightly larger) set of masses 
   contained in the recently published AME2020 database~\cite{Wan21}.  
Finally, to render our results completely reproducible, we provide values 
   for the physical constants as they were used during the readjustment: 
   $\tfrac{\hbar^2}{2 m_n} =\tfrac{\hbar^2}{2 m_p} = 20.73553$ MeV fm$^2$ and 
   $e^2 =1.43996446$ MeV fm, where $e$ is the unit of charge.

\begin{table}[h]
\caption{Infinite nuclear matter properties and final rms $\sigma$ 
         for the BSkG1 parameterization. 
         See Sec.~\ref{Sec:nuc_matter} and Refs.~\cite{Goriely16,Margueron02,Chamel10} for the various definitions.}
\begin{tabular}{l *{1}{d{6.5}} }
\hline
\hline
Properties &      \mc{ BSkG1 }   \\
\hline
$k_F$~[fm]&   1.3280  \\
$n_0$~[fm$^{-3}$]&   0.1582  \\
$a_v$~[MeV]   & -16.088  \\
$J$~[MeV]   & 32.0 \\
$L$~[MeV]   & 51.7  \\
$M_s^*/M$&   0.860  \\
$M_v^*/M$&  0.769  \\
$K_v$~[MeV] &   237.8  \\
$K_{\text{sym}}$~[MeV] &  -156.4  \\
$K^\prime$~[MeV] &   376.7  \\
$G_0$ &   0.35  \\
$G_0^\prime$&  0.98  \\
\hline
\hline
\end{tabular}
\label{tab:prop}
\end{table}

\begin{table}[h]
\caption{Final rms $\sigma$ and mean 
         $\bar \epsilon$ (exp-theory) deviations with respect to the 2457 known 
         masses $M$ where $Z,N \ge 8$ \cite{Wan21}, the 2309 neutron separation energies $S_n$, the 
         2173 $\beta$-decay energies $Q_\beta$, and the 884 measured charge radii $R_c$ \cite{Angeli13} 
         for the mass model generated with the BSkG1 parameterization. 
         The first line gives the model error \cite{Moller88} on all the 2457 measured masses.
         Note that the model was adjusted on the AME2016 database~\cite{Wan17}, but 
         all deviations here are calculated with respect to the 2020 update AME2020~\cite{Wan21}.   }
\begin{tabular}{l *{1}{d{6.5}}  }
\hline
\hline
Results &   \mc{ BSkG1 }  \\
\hline
  $\sigma_{\rm mod}(M)$ [MeV]&   0.734 \\
 $\sigma(M)$ [MeV]&   0.741 \\
$\bar \epsilon (M)$ [MeV] &   -0.026 \\
$\sigma(S_n)$ [MeV]&   0.466 \\
$\bar \epsilon (S_n)$ [MeV] &   0.000 \\
$\sigma(Q_\beta)$ [MeV]&   0.645 \\
$\bar \epsilon (Q_\beta)$ [MeV] &   0.000 \\
$\sigma(R_c)$ [fm] &    0.0239\\
 $\bar \epsilon (R_c)$ [fm] &   -0.0008 \\
\hline
\hline
\end{tabular}
\label{tab:rms}
\end{table}

\subsection{Properties of finite nuclei}
\label{Sec:prop_finit} 

\subsubsection{Nuclear masses and separation energies}

\begin{figure}[]
\centering
\includegraphics[width=.9\linewidth, keepaspectratio]{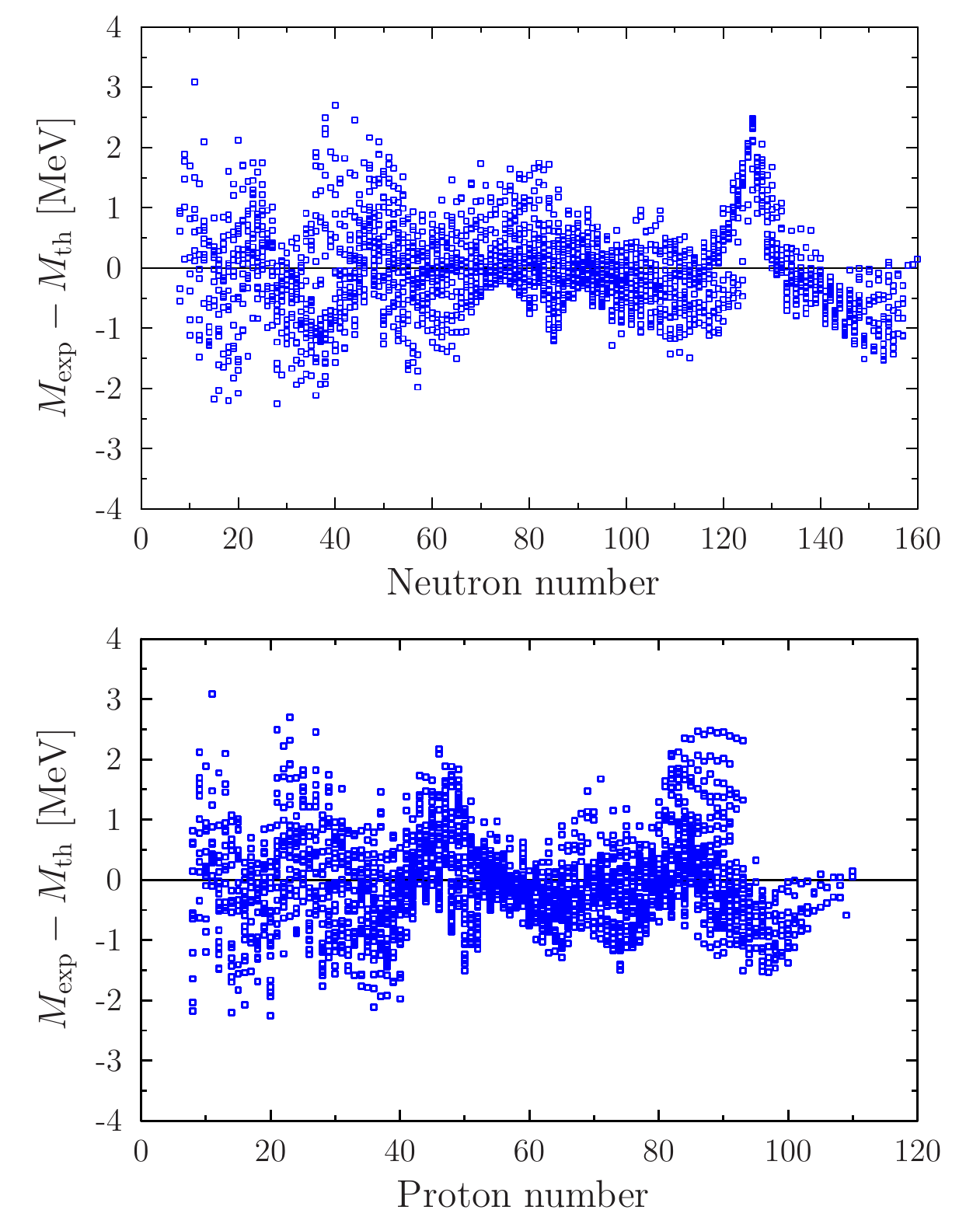}
\caption{(Color online)
    Differences between experimental \cite{Wan21} and theoretical masses as a function 
         of $N$ and $Z$ for BSkG1 (blue squares).
} 
\label{fig:mexp}
\end{figure}

Figure~\ref{fig:mexp} shows the difference between masses calculated with BSkG1 and experimental values as a function of both $N$ and $Z$. The overall 
agreement is excellent and only a limited number of nuclei show a deviation of
more than 2 MeV. The largest systematic deviations can be seen for the 
lightest nuclei, and around the $N=126$ shell closure; smaller deviations
can also be seen in the vicinity of other magic numbers.

The new model achieves an rms deviation on the 2457 known AME2020 masses of 0.741~MeV, 
which is somewhat larger than that achieved by the latest BSk-based HFB mass models~\cite{Goriely16,Goriely13a,goriely2010}. 
In what follows, we will primarily compare to the HFB-21~\cite{goriely2010} and HFB-27~\cite{Goriely13a} 
mass tables, characterized by rms deviations on the nuclear masses of 0.577 and 0.517~MeV
on the AME2020 masses, respectively. The latter is based on a Skyrme EDF of standard form, 
and is quite similar to BSkG1 in many respects, while HFB-21, which has been
used extensively to study r-process abundances (see e.g. Refs.~\cite{Goriely11,Just15,Lemaitre20}),
incorporates non-standard $t_4$ and $t_5$ terms in the Skyrme functional, \textit{i.e.}\ 
terms that are simultaneously momentum and density dependent.

\begin{figure}[]
\centering
\includegraphics[width=\linewidth, keepaspectratio]{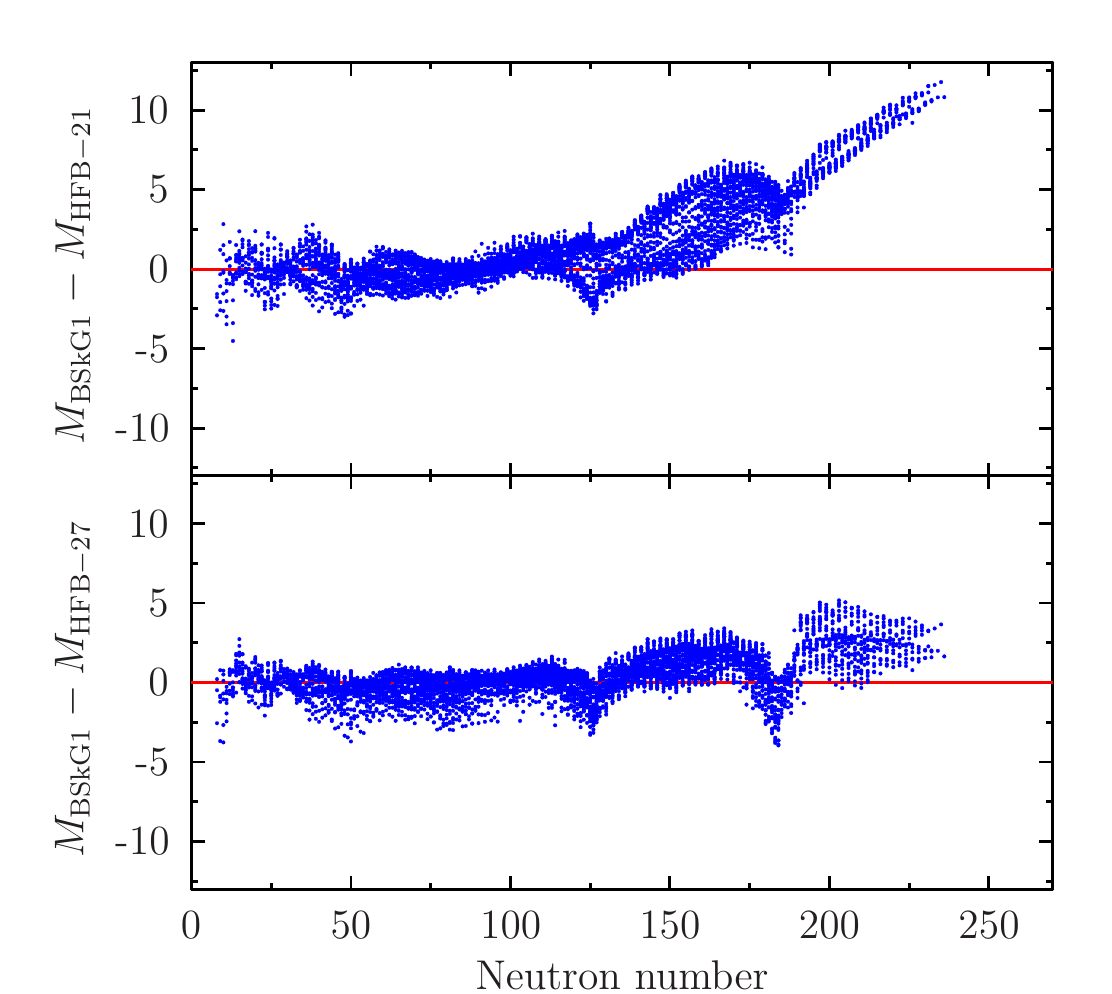}
\caption{ (Color online)
          Comparison between the masses predicted with the BSkG1 interaction and
          HFB-21 (top panel) and HFB-27 (lower panel). All nuclei with $8 \le Z\le 110$ lying between the 
BSkG1 proton and neutron drip lines are included.} 
\label{fig:comp_all}
\end{figure}

Both HFB-21 and HFB-27 achieve an overall lower rms deviation, mainly because of
the following three reasons. The first is their lower value of the symmetry 
coefficient $J= 30$ MeV, which favors the reproduction of 
masses~\cite{Goriely13b} but would lead in our case to an equation of state 
for neutron matter of insufficient stiffness. The second is the treatment of 
the pairing strength: where BSkG1 employs one parameter for each nucleon species, both HFB-21 and HFB-27 employ two different parameters, depending on whether the nucleon number
is even or odd; in addition the BSk pairing cut-off parameter around 16~MeV was found to improve
the mass accuracy with respect to lower values (as used here), at least if no regularization of the pairing
is performed \cite{Goriely06}. Third, both HFB-21 and HFB-27 employ a phenomenological correction 
for collective motion that includes a vibrational component on top of the 
rotational contribution as adopted here. 

Nevertheless, the accuracy of the BSkG1 masses is significantly better than 
those of popular Skyrme parameterizations such as 
SLy4~\cite{Chabanat98} or SLy5s1~\cite{Jodon16}. When evaluated in our framework
for the 561 known masses of even-even nuclei, these parameterizations
give rise to rms deviations of 3.9 and 8.7~MeV, respectively.
The UNEDF collaboration reports rms deviations on the masses of 555 
even-even nuclei of 1.428, 1.912 and 1.950 MeV for the UNEDF0, UNEDF1 and 
UNEDF2 parameterizations, respectively~\cite{Kortelainen14}. The D1M mass model, 
based on the finite-range Gogny interaction \cite{Goriely09} gives an rms 
deviation comparable to that of BSkG1, {\it i.e.}\ 0.810~MeV on the latest 
AME2020 masses. 

When dealing with extrapolated masses away from the experimentally known region,
the BSkG1 masses may differ in a non-negligible way from those predicted by the 
BSk-based HFB mass models. In particular, Fig.~\ref{fig:comp_all} 
   shows, for all $Z\le 110$ nuclei lying between the 
   BSkG1 proton and neutron drip lines,  the mass differences between the BSkG1 and 
   HFB-21(top panel) or HFB-27 (bottom panel) masses as a function of the neutron number. 
   Large differences are observed with respect to HFB-21, especially for the heavy 
   neutron-rich nuclei beyond $N=170$. In contrast, BSkG1 masses are found to be rather similar to HFB-27,  
   with differences typically smaller than 5~MeV. For both HFB models, there
   are, however, significant differences to be seen near $N=184$, and to a lesser 
   extent $N=126$, indicating a different description of shell structure, as discussed in more detail below.

\begin{figure}[]
\centering
\includegraphics[width=\linewidth, keepaspectratio]{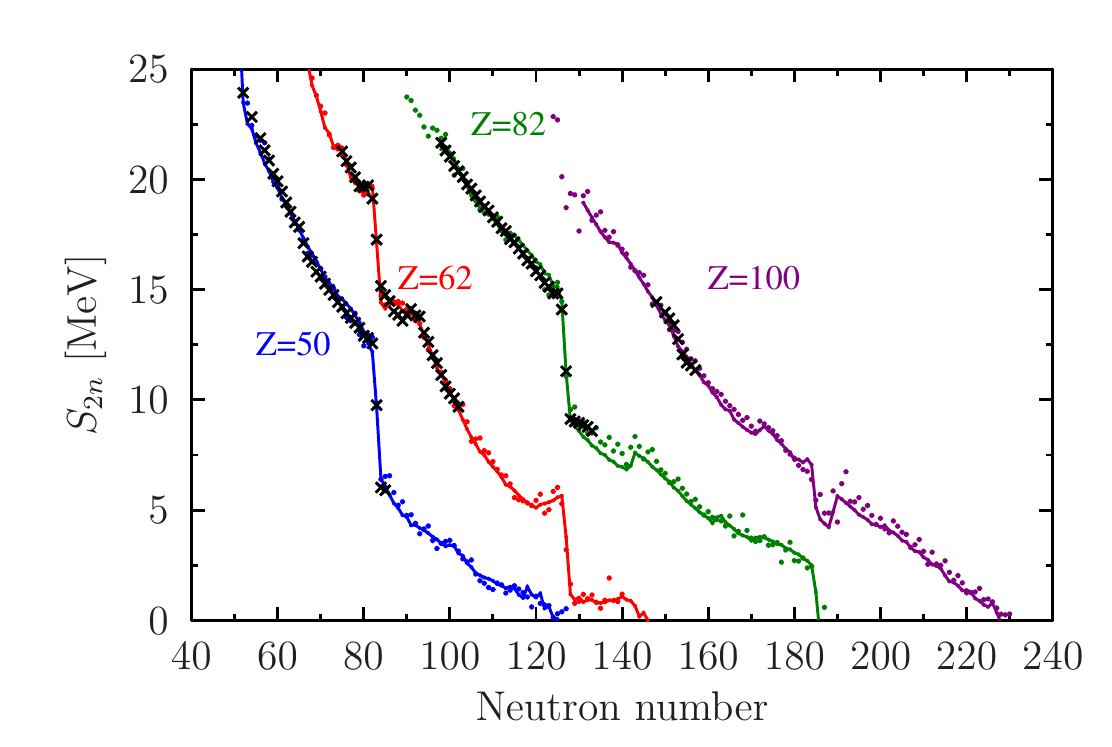}
\caption{ (Color online)
          Two-neutron separation energies $S_{2n}$ for the isotopic chains of Sn, Sm, Pb, and Fm as a function of the neutron number, from AME2020 \cite{Wan21} (black crosses) or predicted either by HFB-21 (dots) or BSkG1 (lines) masses.                
} 
\label{fig:s2n_chains}
\end{figure}

 As typical examples, we show the two-neutron separation energies $S_{2n}$ for the isotopic chains of Sn, Sm, Pb, and Fm in Fig.~\ref{fig:s2n_chains} for both 
   BSkG1 and HFB-21, as well as the available experimental data. 
   Even though differences for individual nuclei can be large, the general
   trends for BSkG1 and HFB-21 are comparable and generally reproduce 
   experimental data rather well. The exceptions are the regions around $N=126$
   and $N=184$, where HFB-21 and BSkG1 each exhibit signs of a different shell 
   structure. Particularly around $N=126$, HFB-21 offers a better description
   of experimental data. Those differences can be seen more clearly in Fig.~\ref{fig:shell_gaps}, as discussed below.
   
 Since separation energies are differences of binding energies, 
   they are directly impacted by the numerical accuracy of the calculation.
   Compared to HFB-21, the coordinate space representation used here for BSkG1 
   results in much smoother separation energies. Irregularities are not 
   totally absent, and can be seen for example near the neutron drip lines 
   for the Sn and Sm isotopes and for neutron-deficient Pb isotopes. Our 
   numerical accuracy in these cases is not limited by the numerical 
   representation, but rather by the resolution of our scan of the 
   $\beta$-$\gamma$ plane     (see Sec.~\ref{sec:semivariational}).

\subsubsection{Shell structure}
\label{sec:shellstructure}

\begin{figure*}[]
\centering
\includegraphics[width=0.8\linewidth, keepaspectratio]{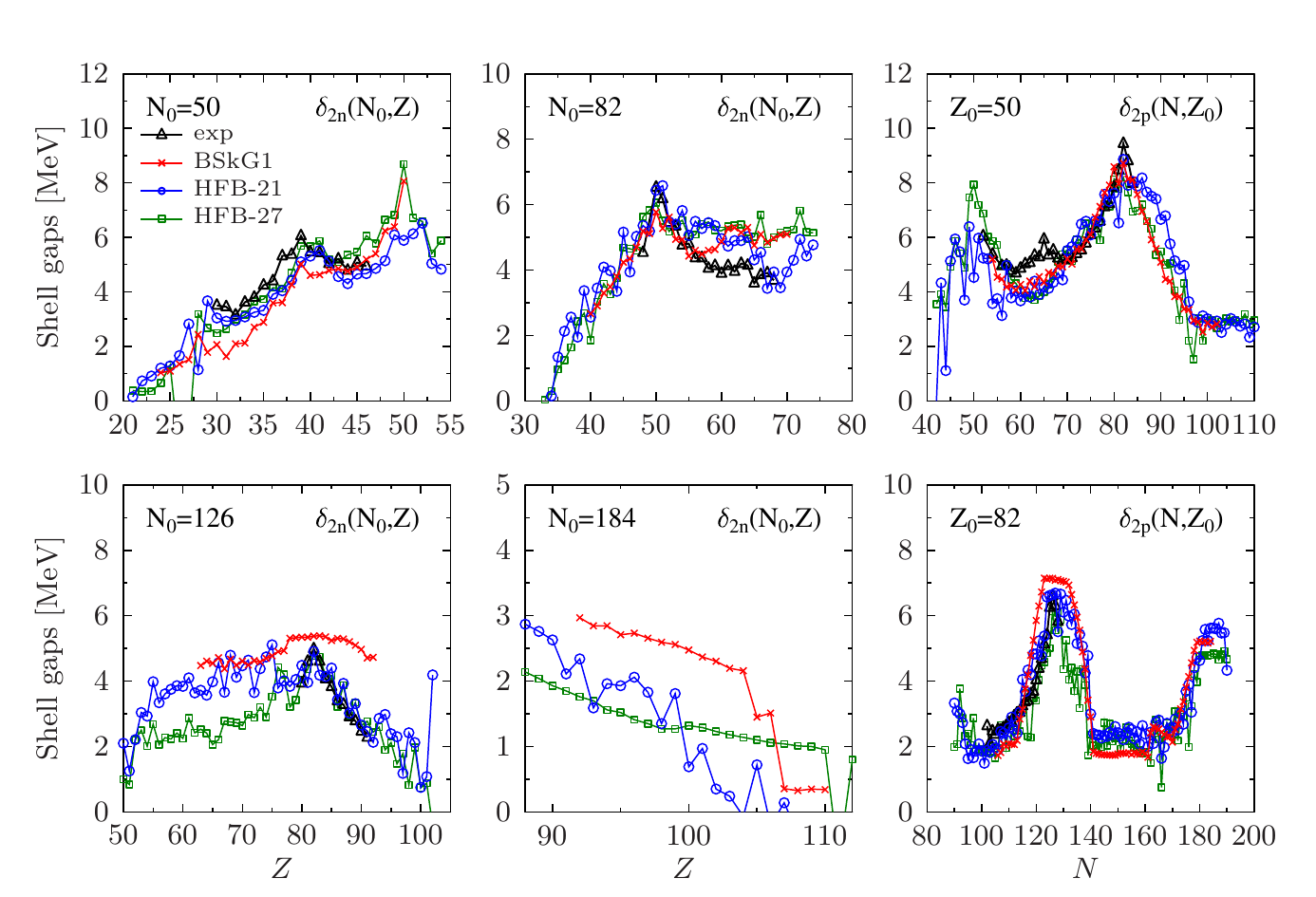}
\caption{(Color online) Shell gaps (see text) as obtained with the BSkG1 
         (red crosses), HFB-21 (blue circles) and HFB-27 (green squares) 
         mass models. Available experimental data \cite{Wan21} are indicated by black triangles. } 
\label{fig:shell_gaps}
\end{figure*}

To investigate BSkG1 shell effects, we consider the usual neutron and 
proton shell gaps $\delta_{2n/p}(N,Z)$, defined as
\begin{subequations}
\begin{align}
\delta_{2n}(N_0, Z) &= S_{2n}(N_0,Z) - S_{2n}(N_0+2,Z) \, ,\\
\delta_{2p}(N  , Z_0) &= S_{2p}(N,Z_0) - S_{2p}(N  ,Z_0+2) \, ,
\end{align}
\label{eq:delta2n2p}
\end{subequations}
where $S_{2n/p}$ are the two-neutron/-proton separation energies. 
We recall that the $\delta_{2n/p}$ serve as an indicator for shell 
closures, but that they are not measures of the size of the gap 
in the single-particle spectrum as they are sensitive to any 
structural change between the three nuclei whose masses enter 
Eq. \eqref{eq:delta2n2p} \cite{Ben02,Ben08}. 
In particular, there is the phenomenon of ``mutually enhanced stability''
that is observed as a peak of the experimental $\delta_{2n/p}$ 
values for doubly-magic nuclei \cite{Zeldes83,Manea20}. In 
mean-field models, it can be partially explained by the onset 
of deformation in adjacent nuclei \cite{Ben02}, but its description 
is  significantly improved when including rotational and vibrational 
corrections of some form \cite{Ben08,Bender06,Delaroche10}, as 
the latter also tend to grow when going away from a doubly-magic 
nucleus. This phenomenon is superposed on the effect of possible 
quenching of the concerned shell closure that would also lead to 
a reduction of the $\delta_{2n/p}$ values when going to weakly-bound 
nuclei \cite{Ben08}.

Figure~\ref{fig:shell_gaps} compares the $\delta_{2n/p}$ values 
across spherical shell closures for the BSkG1, HFB-21 and HFB-27 
sets for the chains of heavy semi-magic nuclei, as well as 
available experimental data. All three mass models produce 
similar values for the $N_0=50$, $82$ and $Z_0=50$, $82$ gaps, 
agreeing with experimental values about equally well.  

The situation is different for the $N_0=126$ and $N_0=184$ shell gaps. 
For the latter, no experimental information is available and the three models 
produce strikingly different predictions. For the $N_0 = 126$ gap, both 
HFB-21 and HFB-27 reproduce the known experimental data rather well, but 
produce different predictions for proton-deficient nuclei. Across the whole
range of proton numbers, however, the BSkG1 model exhibits a structure 
that is qualitatively different from the HFB-21 and HFB-27 models.

\subsubsection{Deformation}

\begin{figure*}[]
\centering
\includegraphics[width=0.49\linewidth, keepaspectratio]{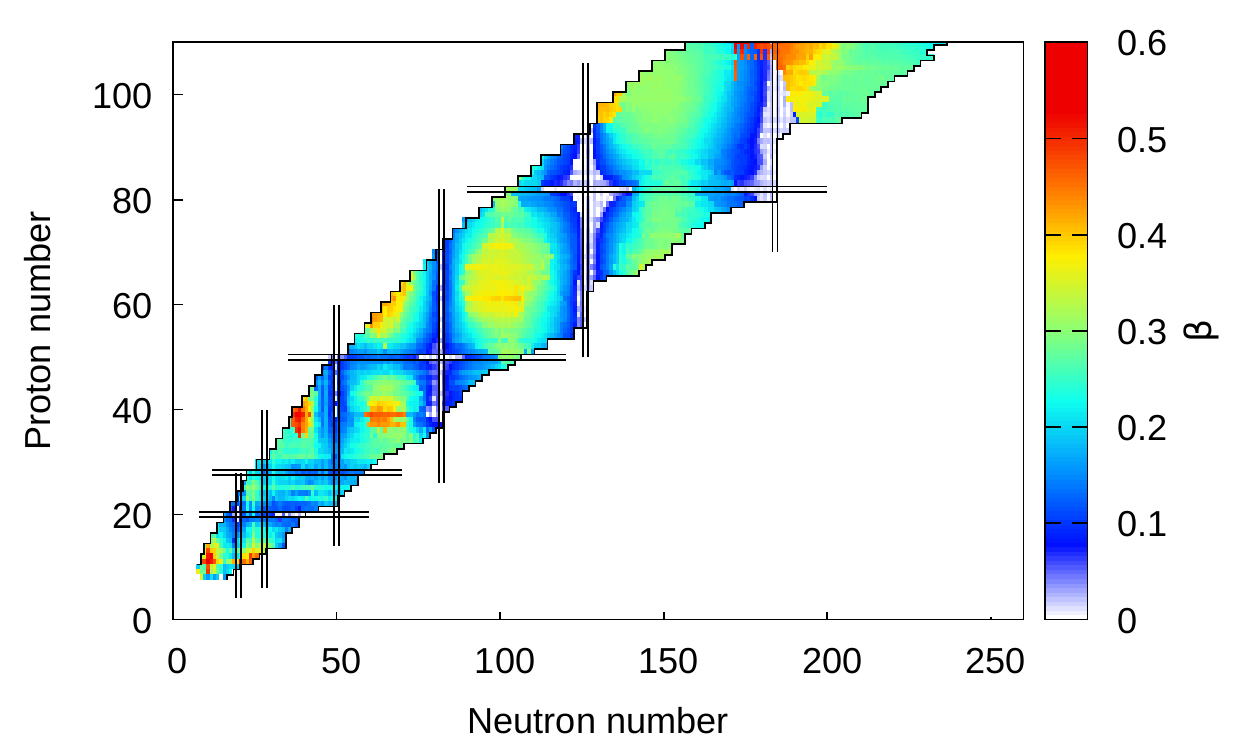}
\includegraphics[width=0.49\linewidth, keepaspectratio]{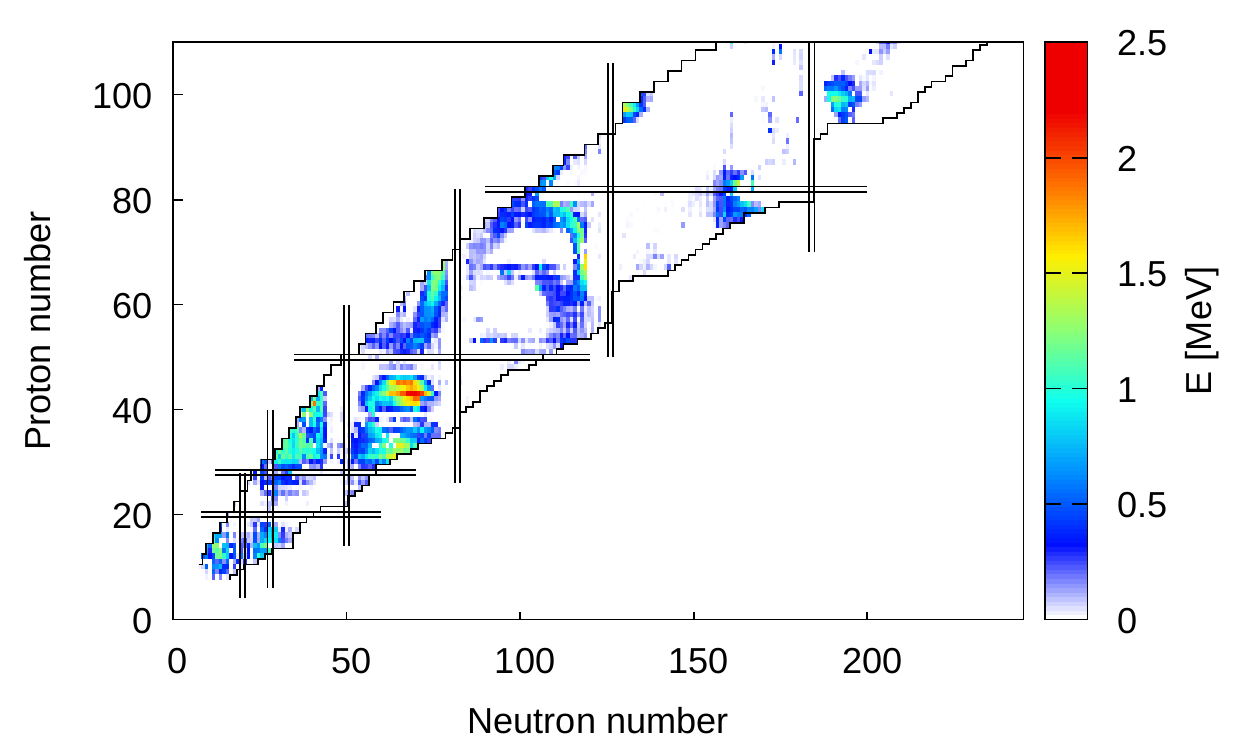}
\includegraphics[width=0.49\linewidth, keepaspectratio]{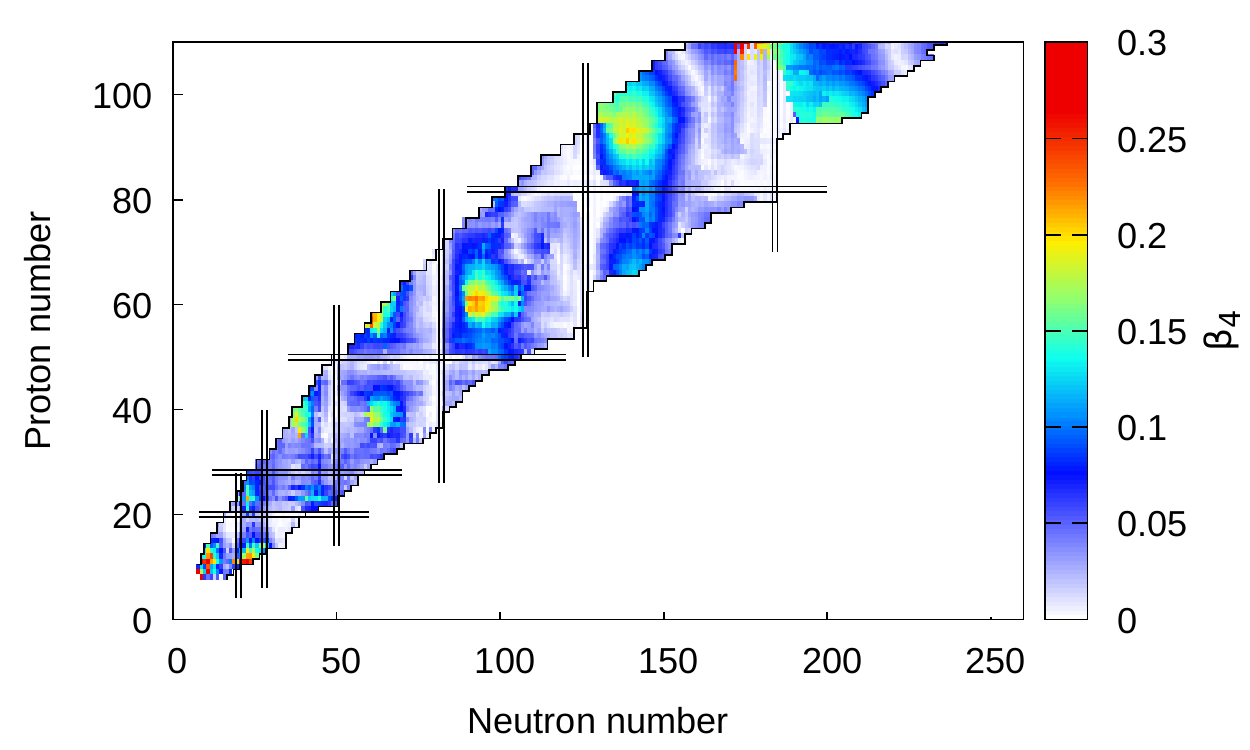}
\includegraphics[width=0.49\linewidth, keepaspectratio]{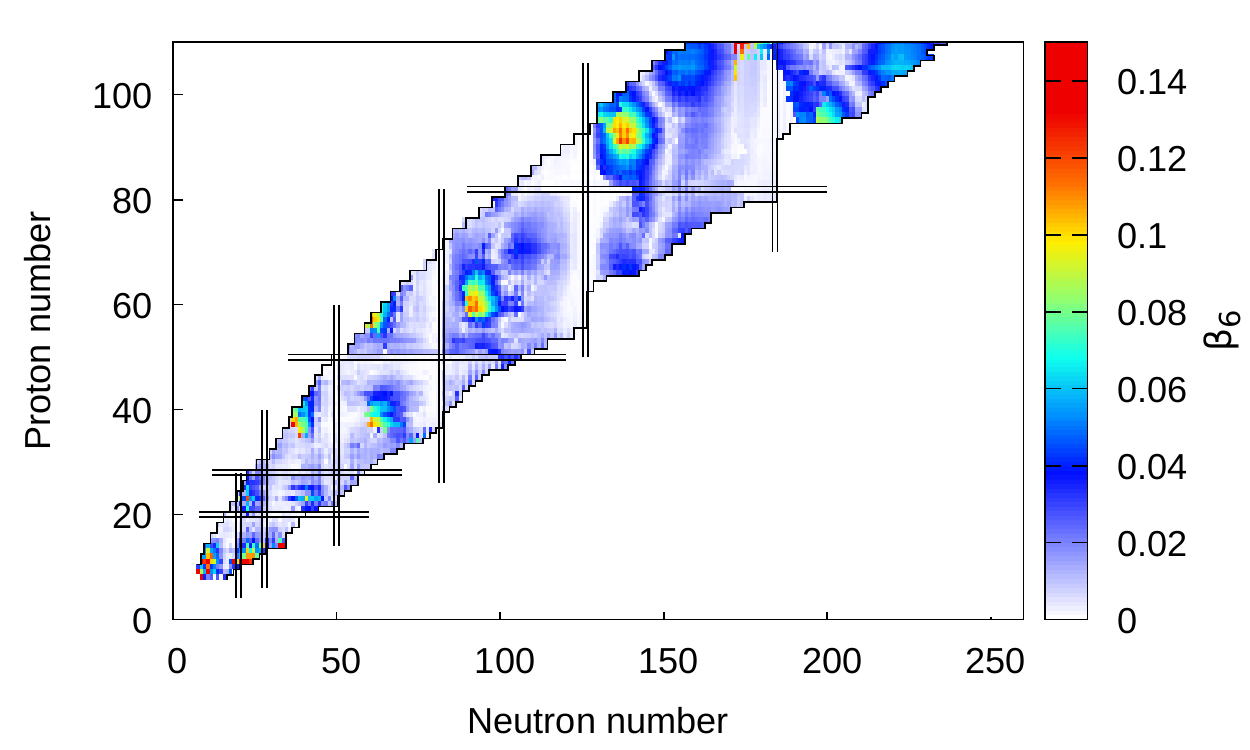}
\caption{(Color online) 
          Top-left panel: Quadrupole deformation $\beta$.
          Top-right panel: Binding energy differences between unrestricted calculations and 
          calculations restricted to axial quadrupole deformations.
          Bottom-left panel: hexadecapole deformation $\beta_4$. 
          Bottom-right: hexacontatetratapole deformation $\beta_6$.
          All quantities extracted from BSkG1 calculations.  }
\label{fig:gain_triax}
\end{figure*}

\paragraph{Global properties} We start by discussing nuclear deformation in the 
BSkG1 model in a global fashion. The top left panel of Fig.~\ref{fig:gain_triax}
shows the quadrupole deformation $\beta$ of Eq.~\eqref{eq:beta} for all 
calculated nuclei. Note that this quantity only indicates the size of the 
quadrupole deformation, and cannot discern between prolate, oblate or triaxial 
shapes. To determine the impact of axial-symmetry breaking on the final outcome, 
we have also performed calculations restricting the nuclei to axial quadrupole 
moments\footnote{Multipole moments of higher order were left unconstrained, such 
that the configurations were not necessarily entirely axially symmetric.}.
The energy differences between these and our unrestricted 
calculations are shown in the top right panel of Fig.~\ref{fig:gain_triax}. We 
see that a large number of nuclei are found to gain from a triaxial deformation 
at least 500~keV in binding energy. This implies that the parameterization makes
significant use of non-axial degrees of freedom to reproduce the known masses: 
restricting the nuclear shape to axial deformations worsens the rms mass 
deviation by more than 100~keV. 

Looking more in detail, we observe several regions where a triaxial deformation 
manifests itself. The largest energy gain, up to 2.5 MeV, is observed for the 
neutron-rich $Z\sim 43$ isotopes. We also find rather deep triaxial minima in 
the neutron deficient $A\sim 70$ region with energy gains on the order of
$1.5$ MeV. Further ``ribbons'' of triaxiality can be seen just below 
the $N=82$ and $N=126$ shell closures, as well as below the $Z=82$ magic 
number. Two intriguing regions are the islands around $Z\sim 96-100$ 
for extremely neutron-deficient and neutron-rich nuclei. Finally, we see that 
several very light nuclei with $Z \leq 20$ also acquire triaxial deformation, 
often due to the rotational correction, as discussed below.

To the best of our knowledge, no similar global survey of triaxial deformation
exists for models based on Skyrme EDFs. We find that regions of 
triaxial deformation are generally centered in the same location as found in 
both microscopic-macroscopic calculations~\cite{Moller06,Moller08}, 
Gogny-HFB calculations based on the D1M mass model~\cite{Goriely09,Hilaire_priv} and  covariant density functional theory \cite{Yang2021}.
The energy gains due to triaxial deformation that we find are generally 
comparable to Gogny results, but significantly larger than those encountered in 
microscopic-macroscopic calculations where energy gains are typically smaller 
than 350~keV and consequently an overall smaller number of triaxial minima is 
found~\cite{Moller06}. More localized studies are more numerous than the (rare) global surveys 
    of triaxiality. Such studies generally concentrate on regions where the 
    BSkG1 model also predicts triaxial ground state deformation, such 
    as the Ge and Se isotopes in the $A \sim 70$ 
    region~\cite{Guo07,Shen11,Niksic14,Bhat14}, the Kr, Sr, Zr, Mo and Ru 
    isotopes in the $A\sim 100$ region~\cite{Bonche85,RodrGuz10,Zhang15,Xiang12,Xiang16,Shi18}, 
    and the neutron-rich rare-earths around $A\sim 190$~\cite{Robledo09,Niksic10,Bhat12}. 

We will compare our nuclear ground state deformations to experimental data below, but we mention here already that triaxial deformation is known
  to also play a role at finite excitation energy: the prime
  example being collective $\gamma$-bands~\cite{BohrMott} that have been 
  documented across the nuclear chart. More exotic examples include chiral 
  bands~\cite{Xiong19}, various kinds of ``wobbling'' 
  bands~\cite{Petrache12,Timar19,Sensharma20} as well as bands based on 
  superdeformed triaxial configurations~\cite{Schnack95,Djongo03}. The calculation
  of such configurations is out of the scope of this paper, but we observe 
  that these phenomena have typically been found in regions of the nuclear 
  chart where the BSkG1 model indicates the importance of triaxial deformation.

Although the nuclear quadrupole moments represent the dominant deformation
   mode, any higher order multipole moments that are unrestricted by symmetry 
   generally take non-zero values as well. For heavy nuclei, we find significant non-zero
   values for $\beta_{\ell}$ up to at least $\ell = 10$, although the deformations 
   naturally become smaller with increasing $\ell$. For example, we show in the 
   bottom panels of Fig.~\ref{fig:gain_triax} the sizes of the hexadecapole 
   ($\ell=4$) and hexacontatetrapole deformations ($\ell=6$) across the nuclear
   chart: there are several regions where the self-consistent optimization utilized these shape degrees of freedom to lower the energy of the nucleus.
   We do not show results for moments beyond $\ell=6$: 
   the associated spherical harmonics oscillate very rapidly and numerically
   calculated integrals involving them are not very precise on a coarsely
   discretized cubic coordinate-space mesh as used here.

\paragraph{Comparison to experiment}

\begin{figure}[]
\centering
\includegraphics[width=0.99\linewidth, keepaspectratio]{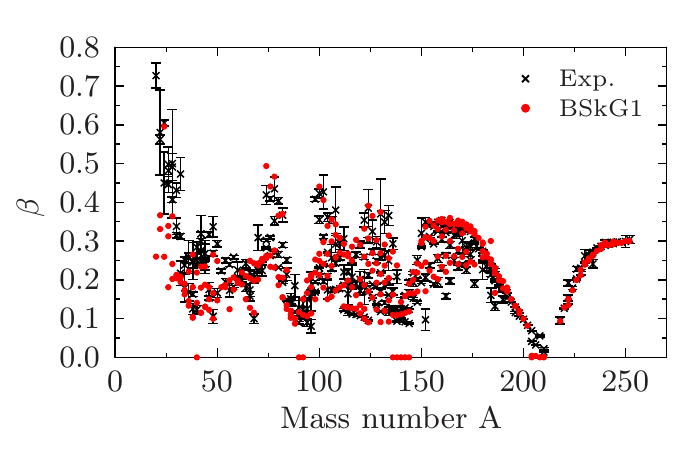} 
\caption{(Color online) Experimental deformations (black crosses) of the 319
                        even-even nuclei of Ref.~\cite{Raman01}, deduced from 
                        the $B(E2)$ transitions to the first excited $2^+$ state.
                        Total quadrupole deformation (red) crosses as calculated
                        with BSkG1.}
\label{fig:comp_raman}
\end{figure}

Most of the experimental information on nuclear deformation concerns the 
    quadru\-pole moments. Figure~\ref{fig:comp_raman} compares the calculated
    quadru\-pole moment $\beta$ for 319 even-even nuclei with the tabulated 
    deformation parameters of Ref.~\cite{Raman01}, extracted from measured 
    $B(E2)$ transition rates. The quadrupole deformation $\beta$ lies roughly 
    between 0.1 and 0.4 for the majority of experimental data points. The 
    BSkG1 model provides a satisfactory description of such well-deformed nuclei,
    and works particularly well for the heavier nuclides beyond $A\sim 160$. For 
    experimental data points with either small ($\beta \leq 0.1$) or 
    for light nuclei with large deformation ($\beta \geq 0.4$), the model 
    performs significantly worse. This is especially visible for the light 
    nuclei below $A \sim 50$ and around $A\sim 208$ near the $N=126$ shell 
    closure. This deficiency is not unexpected, as such nuclei typically cannot 
    be modelled as rotors with static (quadrupole) deformation. In fact, for 
    nuclei with $\beta \leq 0.1$, the excitation spectrum usually indicates that 
    the first $2^+$ state is either vibrational or a multi-quasiparticle 
    excitation, such that the $B(E2, 2^+_1 \to 0^+_1)$ cannot be used to 
    attribute a deformation to the ground state. A more appropriate 
    description of such nuclei would require an improved treatment of collective 
    degrees of freedom, beyond what our description in terms of a single 
    mean-field state (with phenomenological corrections) can provide.

\begin{figure}[]
\centering
\includegraphics[width=0.99\linewidth, keepaspectratio]{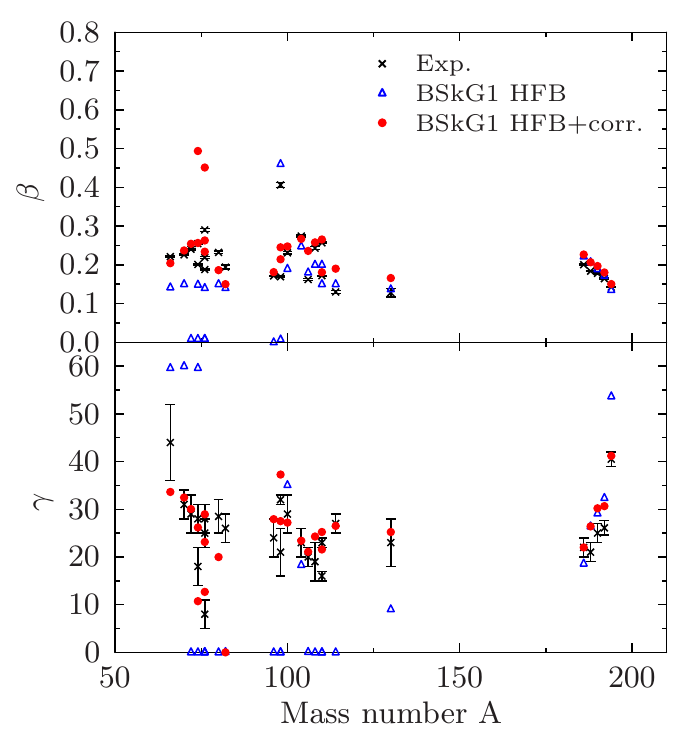} 
\caption{(Color online) 
                        Calculated quadrupole deformation, with (red circles) 
                        and without (blue triangles) the rotational correction,
                        compared to experimental information (black crosses). 
                        Top panel: quadrupole deformation $\beta$ with experimental
                        information from Nudat~\cite{nudat} for
                        $^{66}$Zn~\cite{Rocchini21},
                        $^{70}$Ge~\cite{Sugawara03},
                        $^{72}$Ge~\cite{Ayange16},
                        $^{74}$Ge~\cite{Toh00}, 
                        $^{76}$Ge~\cite{Ayange19},
                        $^{74,76}$Kr~\cite{Clement07},
                        $^{76,80,82}$Se~\cite{Kavka95},
                        $^{98}$Sr~\cite{Clement16},
                        $^{96}$Mo~\cite{ZielinskaPhD},
                        $^{98}$Mo~\cite{Zielinska02},
                        $^{100}$Mo~\cite{Wrzosek12},
                        $^{104}$Ru~\cite{Srebrny06},
                        $^{110}$Cd~\cite{Wrzosek20},
                        $^{114}$Cd~\cite{Fahlander88},
                        $^{106,108,110}$Pd~\cite{Svensson95},
                        $^{130}$Xe~\cite{Morrison20},
                        $^{186,188,190,192}$Os and $^{194}$Pt~\cite{Wu96}.
                        Bottom panel: triaxiality angle $\gamma$, with experimental data points extracted from measured sets of transitional and diagonal $E2$ matrix elements~\cite{Magda_priv} (see text) for the same nuclei.
                        }
\label{fig:comp_exp_def}
\end{figure}

Direct experimental information on triaxial deformation of the nuclear 
    ground state is elusive, but exists for a limited set of nuclei. 
    Through careful analysis of a large 
    number of $E2$ transitions observed in Coulomb excitation experiments, 
    quadrupole rotational invariants of the nuclear ground state can be deduced.
    These invariants can be linked to the deformation of the nucleus in the 
    intrinsic frame~\cite{Kumar72,Cline86}, allowing for the extraction of the mean triaxiality angle $\gamma$. We compare such experimental values of 26 (even-even) nuclei
    for $\gamma$~\cite{Magda_priv} to the values obtained from calculations with
     BSkG1 in the bottom 
    panel of Fig.~\ref{fig:comp_exp_def}. To provide additional context, we also
    compare the calculated size of the quadrupole deformation $\beta$ to values from Nudat calculated from evaluated B($E_2$; 2$^+_1$ $\rightarrow$ 0$^+_1$) transition probabilities \cite{nudat}. The global agreement between experiment and 
    the full BSkG1 model (including the rotational correction) is excellent for 
    both $\beta$ and $\gamma$. We note in particular the reproduction of the
    trend of increasing $\gamma$ and decreasing $\beta$ with mass number in 
    $^{186,188,190,192}$Os and $^{194}$Pt~\cite{Wu96}.

\begin{figure}[]
\includegraphics[width=.99\linewidth,keepaspectratio]{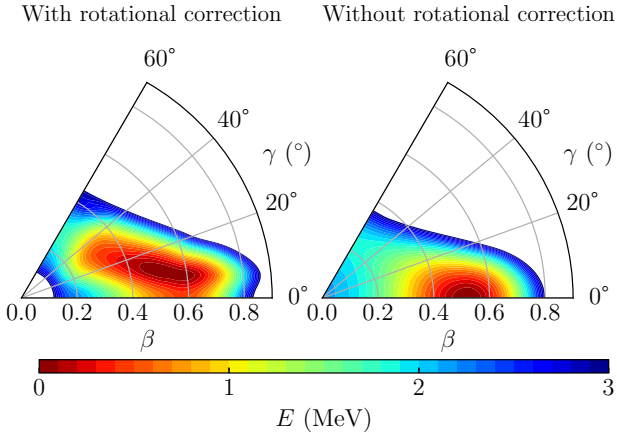}
\caption{ Left panel: deformation energy of $^{24}$Mg relative to its ground state as obtained with 
         BSkG1. Right panel: energy surface with the rotational correction 
         removed, $E_{\rm tot}$ - $E_{\rm rot}$. Both panels are normalized
         to their respective minimum. }
\label{fig:Mg24}
\end{figure}

\paragraph{Impact of the rotational correction}  
The results discussed in the previous paragraphs are influenced significantly 
by the contribution of the rotational correction. This phenomenological 
correction favors (i) larger values of $\beta$ and (ii) triaxial shapes over 
axial ones. Pure mean-field calculations ({\it i.e.} without $E_{\rm corr}$ in
Eq.~\ref{eq:Etot}), would result in a larger number of spherical nuclei, 
smaller overall quadrupole deformations and a smaller number of triaxial
deformations. This is perhaps most noticeable by the appearance of small 
deformations for light semi-magic nuclei in the top left panel of
Fig.~\ref{fig:gain_triax}; in particular, we find no spherical minima for 
nuclei with $N=28$ and/or $Z=28$. This finding, however, does not mean 
that these nuclei are predicted to be static rotors, but rather signals that
these nuclei are so soft that correlation energies are important for their
accurate modelling \cite{Bender06}. Another striking effect of the rotational 
correction is the appearance of large deformation for light nuclei, such as a
minimum at $\beta = 0.590$ for $^{24}$Mg.

Since the rotational correction intends to mimic the impact of symmetry  
restoration, these features of our model emerge naturally, even though they
might be surprising at a first glance. Rotational symmetry-restoration through 
projection techniques~\cite{Bally2021} generally result in lower energies for less symmetric 
configurations, producing slightly deformed minima for nuclei with spherical mean-field 
minima and often very deformed minima for light nuclei~\cite{Bender06}. We 
illustrate the effect of the rotational correction in Fig.~\ref{fig:Mg24}, where
we show the energy of $^{24}$Mg in the $\beta$-$\gamma$ plane. Without 
this correction, the configuration with minimum mean-field energy 
is axially symmetric prolate with $\beta \sim 0.5$. If the correction is included
however, the overall minimum is triaxial with $\beta \sim 0.59$, which is 
comparable to the measured transition charge quadrupole moment $\beta=0.613(14)$ \cite{nudat}.
We see that our example in Fig.~\ref{fig:Mg24} is qualitatively similar to 
the full angular-momentum projected calculations based on the triaxial 
mean-field states of Refs.~\cite{Bender08} and \cite{Rodri10} that are based on 
EDFs of Skyrme and Gogny-type, respectively\footnote{While we cite only 
EDF-based examples, the same effect is present for projected mean-field 
calculations in shell-model valence spaces \cite{Gao15}.}.

Finally, we note that the agreement between COULEX data and 
   the BSkG1 model in Fig.~\ref{fig:comp_exp_def} is at least partially due 
   to the inclusion of the rotational correction. Without it, several of these
   nuclei would exhibit prolate ($\gamma = 0^{\circ}$) or oblate  
   ($\gamma = 60^{\circ}$) minima, generally at a lower value of $\beta$ than
   experimental data indicates. We take this as an indication that the modelling 
   of these nuclei in terms of a pure mean-field state is insufficient and 
   beyond-mean-field effects are important, even if they are only schematically
   included as we do here.

\begin{figure}[]
\includegraphics[width=\linewidth, keepaspectratio]{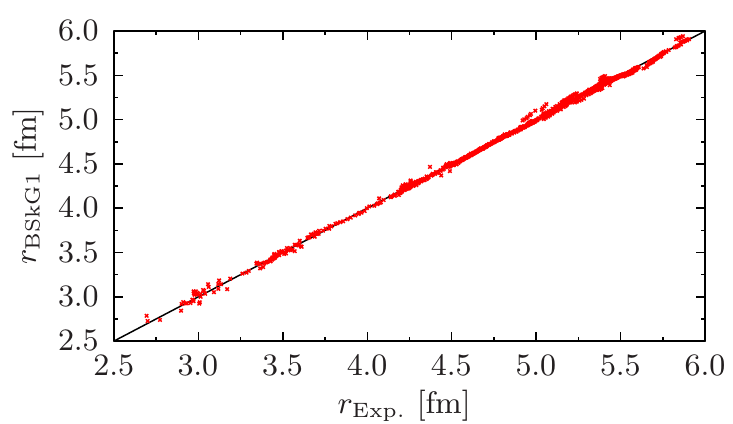}
\includegraphics[width=\linewidth, keepaspectratio]{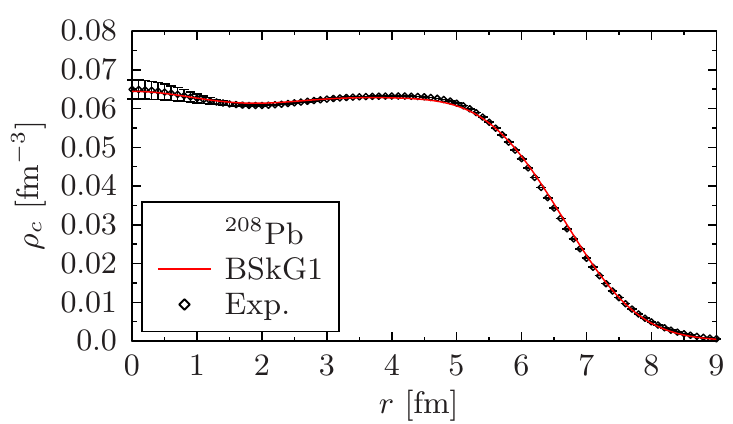}
\caption{(Color online) 
        Top panel: comparison of BSkG1 charge radii to experimental values
         taken from Ref.~\cite{Angeli13}. Lower panel: calculated radial 
         charge density of $^{208}$Pb compared to experimental data 
         of Ref.~\cite{Euteneuer78}.}
\label{fig:chargeradii}
\end{figure}

\subsubsection{Charge distribution}

Charge radii are obtained with an accuracy similar to those 
of the BSk mass models, \textit{i.e.}\ with an rms deviation of 0.024~fm with 
respect to the 884 measured values~\cite{Angeli13}. We show the agreement 
between known and calculated charge radii in the top panel of 
Fig.~\ref{fig:chargeradii}. We recall that the charge densities are obtained
through folding of the neutron and proton point densities with appropriate 
form factors (Sec.~\ref{sec:Coulomb}).

As a further point of comparison, we show in the 
bottom panel of Fig.~\ref{fig:chargeradii} the calculated charge distribution 
of $^{208}$Pb, to be compared to the measured values of Ref.~\cite{Euteneuer78}.
The overall agreement is excellent, a quality shared with the 
BSk models~\cite{Goriely13b}. It should also be emphasized that we obtain a  
neutron skin thickness for $^{208}$Pb of 0.18~fm, as expected from the symmetry 
energy constrained to $J=32$ MeV~\cite{Goriely13b}.

\subsubsection{Moments of inertia and pairing properties}

\begin{figure}[t]
\centering
\includegraphics[width=\linewidth, keepaspectratio]{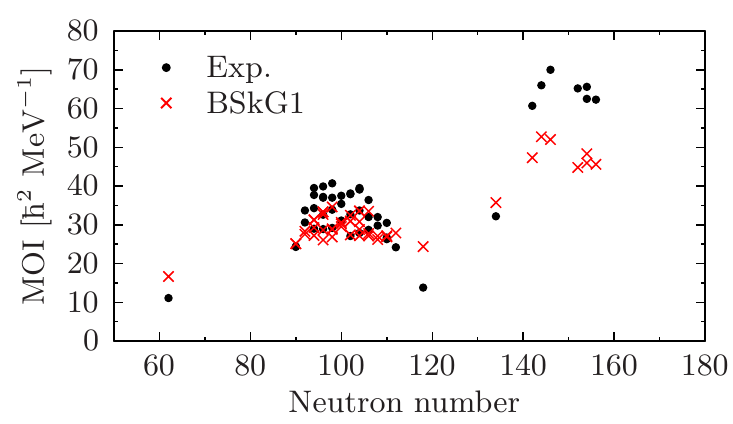}
\caption{Belyaev MOI obtained with BSkG1
        (red crosses) compared to the experimental data (black dots) 
        compiled in Refs.~\cite{Zeng94,Afanasjev2000,Pearson1991} . } 
\label{fig:MOI}
\end{figure}

Figure~\ref{fig:MOI} compares the calculated MOI to the 
48 experimental values included in the objective function, as discussed in 
Sec.~\ref{sec:ingredients}. The calculated MOIs systematically
underestimate the experimental data by about 10\%, particularly for the heavy $N\sim 150$
isotopes, but provide an overall acceptable description of the rotational
properties of these even-even nuclei\footnote{The comparison of the MOI of triaxial configurations with experiment is not trivial, see 
\ref{app:rotational}.}. In addition, we remark that it is natural that 
our calculations somewhat underestimate the experimental values; if we would 
consider the Thouless-Valatin MOI instead of the simple Belyaev MOI, the 
calculated values would increase by roughly 30\%~\cite{Petrik18}.

\begin{figure}[t]
\centering
\includegraphics[width=\linewidth, keepaspectratio]{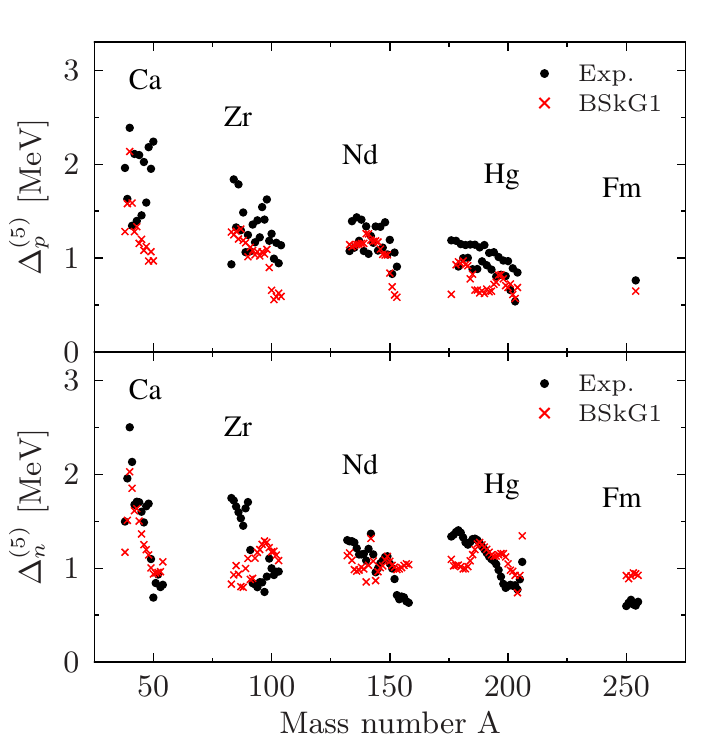}
\caption{(Color online) Comparison between the five-point proton (top) and neutron (bottom) 
          mass differences $\Delta^{(5)}_{n,p}$ for isotopes of Ca, Zr, Nd, Hg, and Fm, 
          as obtained from experiment (black dots) and BSkG1 (red crosses). }
\label{fig:gap_5points}
\end{figure}

To estimate the BSkG1 pairing effects, we compare the calculated five-point mass differences in 
Fig.~\ref{fig:gap_5points} with experimental values for a representative
sample of nuclei, namely the isotopic chains of Ca, Zr, Nd, Hg, and Fm. 
The overall size of the mass differences, for both protons and neutrons, is 
reasonably well described.

The overall agreement with experiment of the MOI and the mass differences 
in Figs.~\ref{fig:MOI} and \ref{fig:gap_5points} is a direct consequence of the 
inclusion of the former into the objective function (Eq.~\ref{eq:sigmadef}).
Without such a constraint, the model readjustment to only the absolute nuclear
masses would have resulted in an overall lower rms on the known masses, 
$\sigma(M) \sim 0.66$~MeV, but with unrealistically large odd-even staggering 
and unrealistically small MOIs. At the other extreme, readjusting the 
pairing properties only on the odd-even staggering of the masses would 
result in severe underbinding for the majority of open-shell nuclei 
with small or vanishing deformations, mostly due to the absence of correlations 
beyond the rotational correction~\cite{Goriely06}. An increase of the pairing
strength would then help the fit to emulate the missing collective binding 
energy for such nuclei in a global fashion, while deteriorating the description 
of the local, more fine-grained, odd-even staggering effect. 
This would have impacted negatively not only the description of other quantities sensitive
to the pairing strength, such as the nuclear level densities, fission barriers
and the MOIs discussed here, but also our confidence in extrapolation 
of the results to exotic nuclei. 

\subsection{Nuclear matter properties}
\label{Sec:nuc_matter} 

The main properties of infinite charge-symmetric or pure-neutron matter are 
listed in Table~\ref{tab:prop}. As discussed in Sec.~\ref{sec:ingredients}, some of these
properties have been constrained to a given value or a restricted range during the fitting procedure
in order to reproduce experimentally extracted information or predictions from ab-initio calculations, as discussed in Ref.~\cite{Goriely16}.
These properties concern in particular the symmetry coefficient $J=32$~MeV, the compressibility
$K_v=237.8$~MeV through the $\gamma=0.3$ parameter of the Skyrme EDF of Eq.~\eqref{eq:Skyrme}\cite{Chabanat97}, and the isoscalar effective mass $M^*_s/M=0.86$.
The Fermi wave number $k_F$ is known to drastically affect the nuclear radii and for this reason has been
adjusted to minimize the overall mean deviation between experimental and predicted charge radii. The other
properties of infinite nuclear matter directly result from the adjustment procedure and are found
to be in relatively good agreement with values extracted from measurements or ab-initio calculations, as discussed below.

\begin{figure}
\centering
\includegraphics[width=\linewidth, keepaspectratio]{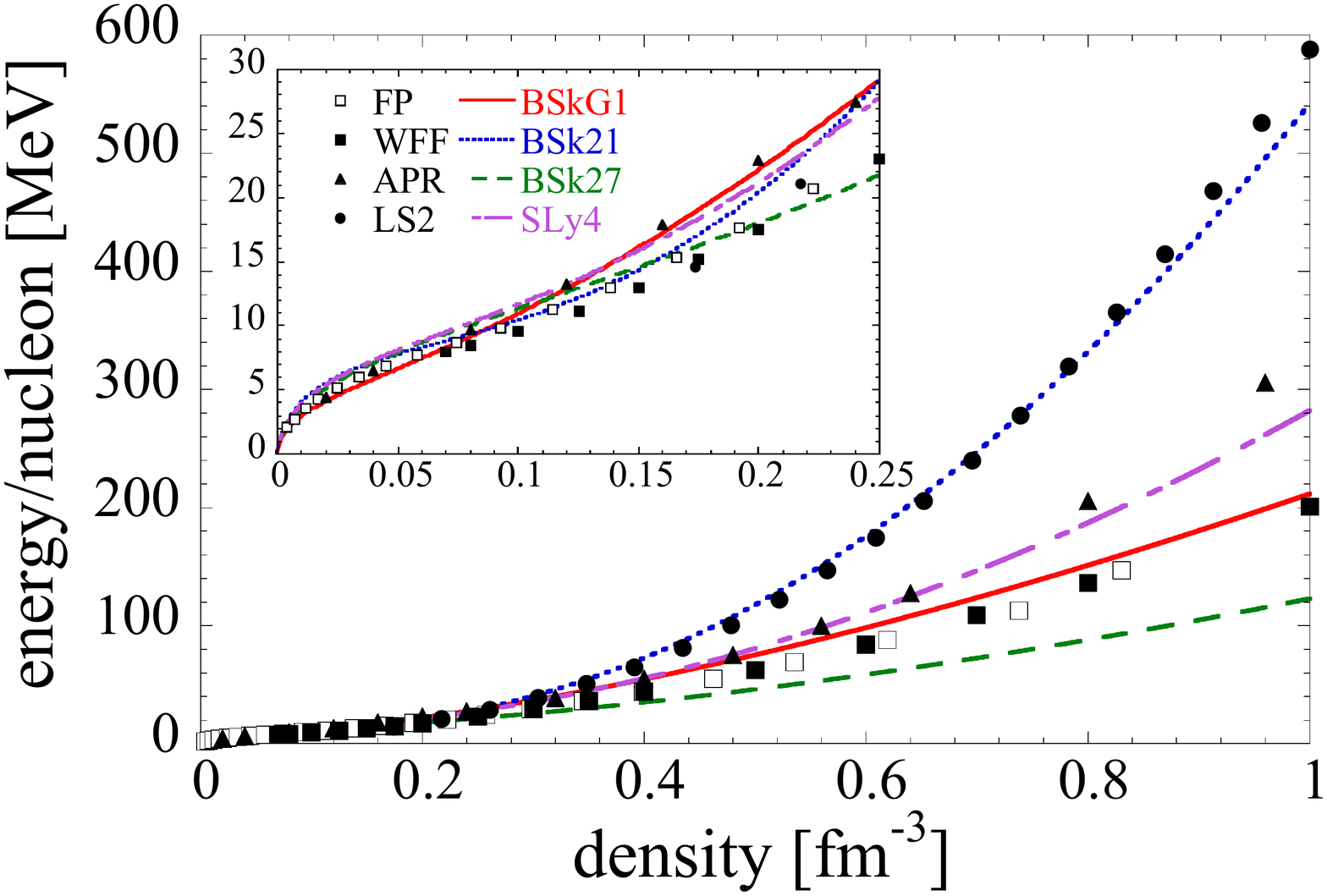}
\caption{(Color online) Zero-temperature equations of state for pure neutron matter with BSkG1, compared to ab-initio
calculations referred to as FP, WFF, APR and LS2, as well as the 
BSk21~\cite{goriely2010}, BSk27~\cite{Goriely13a} and SLy4 \cite{Chabanat98} 
Skyrme parameterizations. FP corresponds to the ab-initio
calculation of Friedman \& Pandharipande \cite{Friedman81}, 
WFF to ``UV14 plus TNI'' of Ref.~\cite{Wiringa88}, 
APR of ``A18 + $\delta\,v$ + UIX$^*$" in Ref.~\cite{Akmal98} and
LS2 to V18 in Ref.~\cite{Li08}. The upper left insert is a zoom in view of densities below 0.25~fm$^{-3}$.} 
\label{fig:neutmat}
\end{figure}

In Fig.~\ref{fig:neutmat} we show the equation of state for pure 
neutron matter calculated with  BSkG1. Despite the (imposed) value of the symmetry 
coefficient $J$, the equation of state of neutron matter remains rather soft at high 
densities with respect to the ab-initio calculations of APR \cite{Akmal98} and 
LS2 \cite{Li08}, but is in agreement with FP \cite{Friedman81} and 
WFF \cite{Wiringa88}. Compared to the BSk21 and BSk27 interactions, 
BSkG1 is somewhat intermediate in stiffness, and comparable to the SLy4
interaction, also characterized with $J=32$~MeV. The maximum mass of non-rotating NSs
for the BSkG1 equation of state is estimated to reach 1.79~$M_\odot$, assuming the NSs are in $\beta$-equilibrium
at zero temperature. This value is certainly below the observed limit of $2.08\pm0.07 M_\odot$ for  pulsar PSR J0740+66220 \cite{Fonseca21}.
Higher values of the symmetry coefficient $J$ are not favored either by mass fits \cite{Goriely16}, or by ab-initio calculations of infinite neutron matter at low density (see the tendency for BSkG1 to underestimate ab-initio calculations in the insert of Fig.~\ref{fig:neutmat}). A compatible stiffer equation of state, and hence higher NS masses, can be obtained by including density-dependent $t_4$ and $t_5$ terms in the EDF, 
as found with the BSk interactions starting from BSk18 \cite{Chamel09}. Also note that the symmetry coefficient $J=32$~MeV and slope $L=51.7$~MeV are found to be fully compatible with the experimental constraints from heavy-ion collisions \cite{Tsang09}, neutron-skin thickness in Sn isotopes \cite{Chen10} and the analysis of the giant dipole resonance \cite{Lattimer13,Trippa08}, as summarized in Refs.~\cite{Goriely16,Lattimer13,Fortin16}.

\begin{figure}
\centering
\includegraphics[width=\linewidth, keepaspectratio]{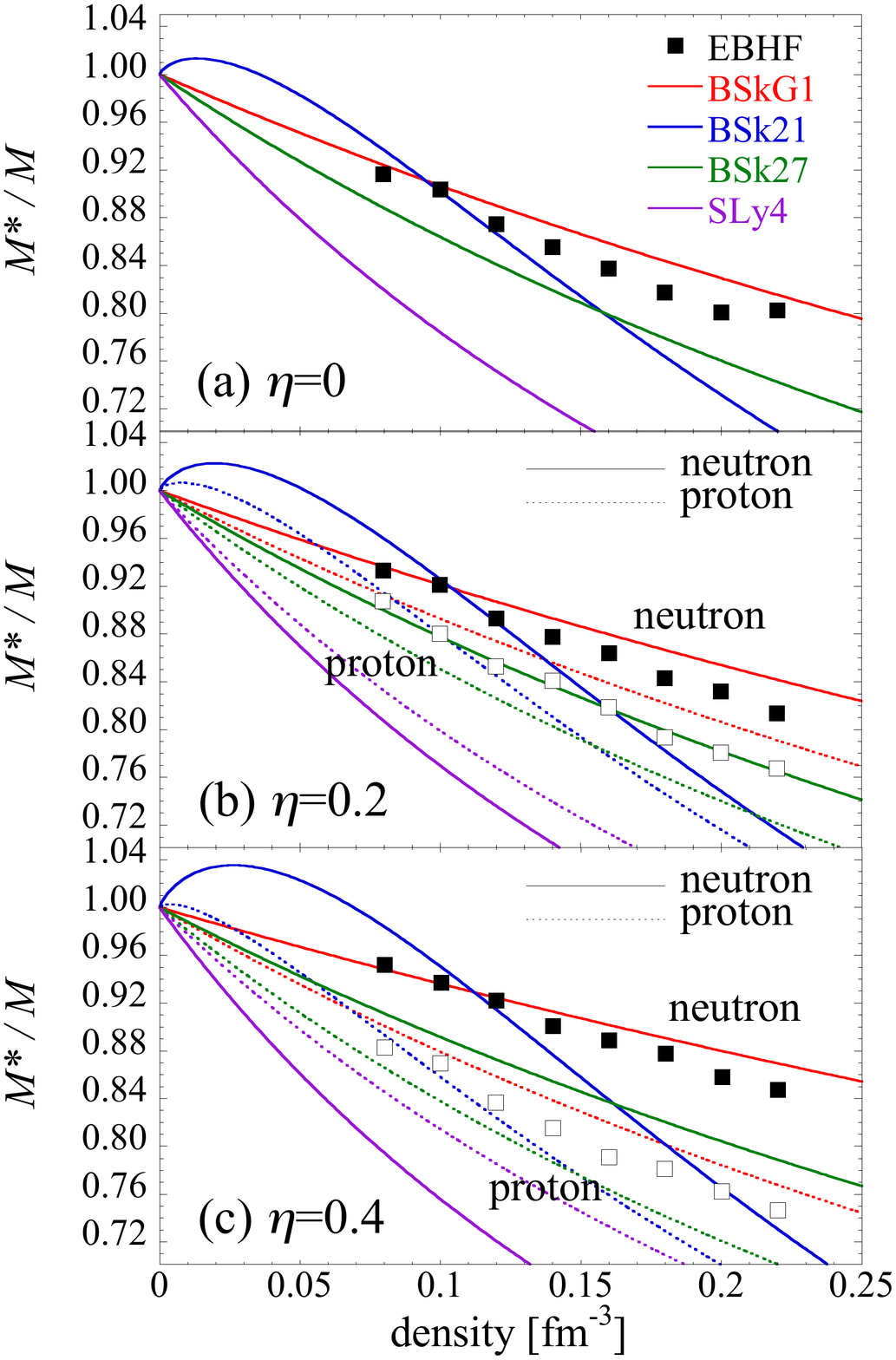}
\caption{(Color online) Neutron and proton effective masses in symmetric nuclear
matter (a) as well as in asymmetric matter with an asymmetry $\eta=0.2$
(b) and 0.4 (c) obtained with the BSkG1 (red lines), BSk21 (blue
lines), BSk27 (green lines) and SLy4 (purple lines). For asymmetric matter, the
neutron effective masses are shown with solid lines and the proton ones with 
dashed lines. The EBHF calculations of Ref.~\cite{Cao06b} are
shown by squares for comparison (full and open squares correspond to the neutron and proton effective masses, respectively).} 
\label{fig:meff}
\end{figure}

As shown in Table \ref{tab:prop}, the fit led to an isoscalar effective mass at the saturation density of $M^*_s/M=0.86$ 
in good agreement with the values obtained by the EBHF calculations of Ref.~\cite{Cao06,Zuo02}. 
The isovector effective mass, $M^*_v/M$, that emerged from the fit is found to be lower than the isoscalar effective mass, 
which implies that the neutron effective mass is larger than the proton effective mass in neutron-rich matter.
This mass hierarchy between neutrons and protons  is consistent with measurements of the 
isovector giant dipole resonance \cite{Lesinski06}, and has been confirmed in ab-initio many-body calculations \cite{Cao06,Cao06b}. 
In particular, for BSkG1, the magnitude of the splitting $M^*_s-M^*_v=0.091$ is in good agreement with such ab-initio estimates of 0.098.
A similar splitting was found with BSk21 (0.09) and BSk27 (0.08), in contrast to SLy4 which is characterized by a negative splitting
of $-0.11$, {\it i.e.}\ an opposite mass hierarchy $M^*_p>M^*_n$. These splittings are also illustrated in Fig.~\ref{fig:meff}
which compares, as a function of the density $n$, neutron and proton effective masses in symmetric nuclear
matter ($\eta=(n_n-n_p)/(n_n+n_p)=0$) as well as in asymmetric matter with $\eta=0.2$ and 0.4 obtained with the BSkG1,
BSk21, BSk27, SLy4 interactions and the EBHF calculation of Ref.~\cite{Cao06b}. Most of the BSk forces, like BSk21 and BSk27, have an isoscalar effective mass at the saturation density constrained to 0.80, except BSk30-32 which adopted a higher value of 0.84.
Note that the BSk21 non-linearity of $1/M^*_n$ and $1/M^*_p$ with density is due to the terms in $t_4$ and $t_5$.

\begin{figure*}[]
\centering
\includegraphics[width=0.8\linewidth, keepaspectratio]{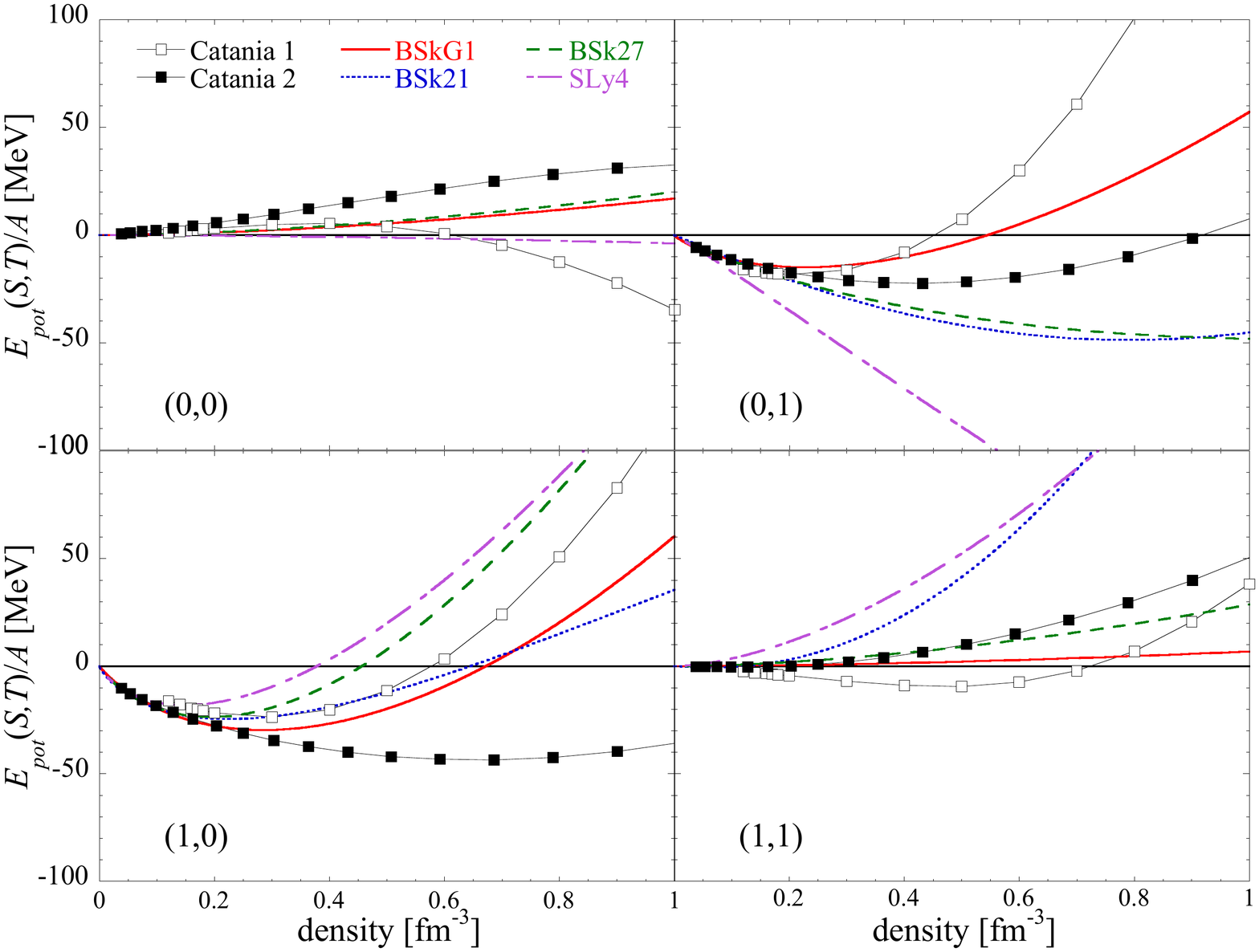}
\caption{ (Color online) 
         Potential energy per particle $E_{\rm pot}/A$ in each $(S,T)$ channel as
         indicated as a function of density for charge-symmetric infinite 
         nuclear matter for BSkG1, BSk21~\cite{goriely2010}, 
         BSk27~\cite{Goriely13a} and SLy4 \cite{Chabanat98}. The open and
         solid squares correspond to the ``Catania~1" \cite{Li08} and 
         ``Catania~2" BHF calculations \cite{Zhou04}, respectively.        
} 
\label{fig:est}
\end{figure*}

Fitting our forces to the mass data is not a sufficient condition  for ensuring 
a realistic distribution of the potential energy per nucleon  among the four 
two-body spin-isospin $(S, T)$ channels in charge-symmetric infinite nuclear 
matter. Figure~\ref{fig:est} shows this distribution for our new interaction as a 
function of the density  and compares it with two different Brueckner-Hartree-Fock (BHF)
calculations labeled ``Catania 1" \cite{Li08} and ``Catania 2" \cite{Zhou04}.
For BSkG1, as well as BSk21, BSk27 and SLy4, the energies are calculated using 
the expression from Ref.~\cite{Lesinski06}, setting the coupling 
constants $C^{sT}_t$ to zero for consistency with the choices 
made for the energy density of Eq.~\eqref{eq:Skyrme} \cite{Ben03,Chamel10}.
 Given the  evident uncertainty in what the real distribution actually is, the level of 
agreement we have found with our new BSkG1 interaction can be regarded as satisfactory, and significantly better than that obtained with the SLy4 functional, in particular for the $T=1$ channels.

\subsection{Application to the r-process nucleosynthesis}
\label{Sec:r_proc} 
As an application of the 
new mass model, abundance distributions
resulting from the r-process nucleosynthesis in NS mergers have 
been calculated with BSkG1 masses. The  neutron capture and photoneutron 
astrophysical rates have been calculated for all nuclei with 
$8 \le Z \le 110$ lying between the proton and neutron drip lines on the basis 
of the ground state properties obtained with the BSkG1 mass model.
Both the dynamical and disk ejecta of a NS binary system have been considered \cite{Goriely11,Just15}, 
as detailed below.

\begin{figure}[h]
\centering
\includegraphics[width=1.05\linewidth, keepaspectratio]{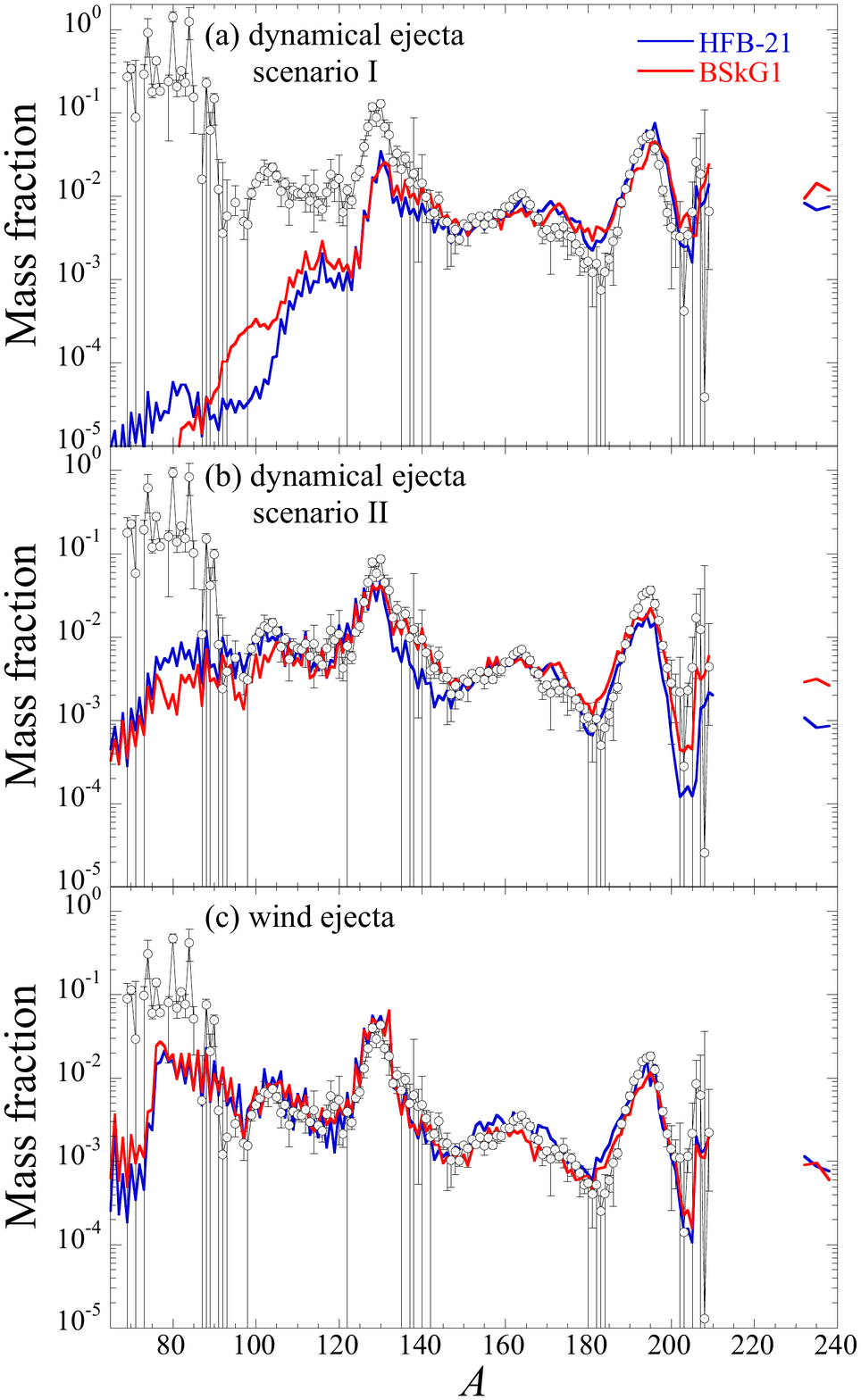}
\caption{(Color online) (a) Mass fraction of the $4.9\times 10^{-3}~M_{\odot}$ of material ejected in scenario I of 
         the 1.365--1.365$M_{\odot}$ NS-NS merger model as a function of the 
         atomic mass $A$. The red curve obtained with the BSkG1
         nuclear masses and corresponding neutron capture and photoneutron rates is compared to the results obtained with HFB-21 \cite{goriely2010} (blue curve). 
         The solar system r-abundance distribution with corresponding error bars (open circles), 
         arbitrarily normalized to the HFB-21 predictions in the $A\simeq 164$ region, is shown for comparison \cite{Goriely99}.
          (b) Same as (a) for scenario II of the dynamical ejecta.
          (c) Composition of the $2.5 \times 10^{-2}~M_{\odot}$ of material ejected from the post-merger BH-torus remnant characterized by a torus mass  of $0.1 M_{\odot}$ and a 3\,$M_\odot$ BH~\cite{Just15}.} 
\label{fig:rpro}
\end{figure}

In the case of the dynamical ejecta, the same model as described in Ref.~\cite{Lemaitre20} and 
corresponding to a symmetric 1.365--1.365$M_{\odot}$  binary system obtained with the SFHo equation of state \cite{Steiner2013}
is adopted. Like in Ref.~\cite{Lemaitre20}, we consider two distinct scenarios reflecting the possible impact (or not) 
of neutrino absorption on the initial neutron-richness of the dynamical ejecta. In scenario I, all weak interactions on free nucleons
are neglected and the mean initial electron fraction at the time the temperature has dropped below 10~GK corresponds
to $\langle Y_e \rangle=0.03$. This case is expected to mimic the nucleosynthesis from NS-black hole (BH) mergers or 
the prompt collapse of mass-asymmetric NS-NS mergers. In the second case (scenario II), weak nucleonic interactions
are incorporated following the parametric approach of Ref.~\cite{Goriely15}, in terms of prescribed neutrino luminosities and mean energies, 
but guided by the hydrodynamical simulations of Ref.~\cite{Ardevol19} which includes a self-consistent approximated treatment of 
neutrino emission and absorption. This scenario is characterized with $\langle Y_e \rangle=0.23$ and consequently gives rise to
an r-process nucleosynthesis less efficient than in scenario I.
Details about the nucleosynthesis calculations,
additional nuclear inputs and the astrophysical scenario can be found in Ref.~\cite{Lemaitre20}. For both scenarios, the abundance 
distribution of the ejected material obtained with BSkG1 reaction rates
is shown in Fig.~\ref{fig:rpro} and compared to those found with the HFB-21 
nuclear inputs \cite{goriely2010}, as used in Ref.~\cite{Lemaitre20}.
Note that in both simulations the same  $\beta$-decay rates as well as $\beta$-delayed neutron emission probabilities are taken from the relativistic mean-field model of Ref.~\cite{Marketin16}.  
Both mass models are found to give rise to abundance distributions that reproduce well the r-process peaks and the rare-earth bump. The BSkG1 masses tend to produce more nuclei around $A \simeq 140$, as well 
as Pb-group elements and actinides, essentially due to its stronger shell effect around  $N=184$ (as discussed in Sec.~\ref{sec:shellstructure}) which causes a larger accumulation of nuclei along the super-heavy $N=184$ bottleneck during the neutron irradiation.

In addition to the dynamical ejecta, the neutrino and viscously driven outflows generated during the post-merger remnant evolution of the relic BH-torus system can be expected to give rise to a significant ejection of $A \ge 80$ r-process-rich material \cite{Just15}. Full details about the hydrodynamical simulations can be found in Ref.~\cite{Just15}. We consider here a representative sample of 2075 trajectories ejected from a system characterized by a torus mass of   $M_{\mathrm{torus}}=0.1 M_{\odot}$\ and a 3\,$M_\odot$ BH (corresponding to the M3A8m1a5 model of Ref.~\cite{Just15}). The total mass ejected from the BH-torus system amounts to $2.5 \times 10^{-2}~M_{\odot}$ and the outflow is characterized by a mean initial electron fraction $\langle Y_e \rangle=0.24$.

The impact of the new masses on the composition of the BH-torus disk ejecta is shown in Fig.~\ref{fig:rpro}c where both mass models are seen to give rise to rather similar abundance distributions, especially around the second and third r-process peaks. BSkG1 masses tend to produce slightly less rare-earth elements than HFB-21.
 
\section{Conclusions and outlook}
\label{Sec:Summary} 

\subsection{Conclusions}
We have presented the BSkG1 mass model, based on an energy density functional 
of the Skyrme type. The BSkG1 interaction has been adjusted on essentially all known 
nuclear masses, rendering the model well-suited for nuclear applications. 
The model achieves an rms deviation of 0.741~MeV on the 2457 known masses 
of the AME2020 database \cite{Wan21} and an rms deviation of 0.0239~fm on the 884 charge 
radii from Ref.~\cite{Angeli13}. To obtain realistic pairing strengths, we have
included experimental information on the moment of inertia of a set of heavy 
nuclei in the adjustment procedure, resulting in a fair description of the 
pairing and rotational properties across the nuclear chart. 
We have in addition shown that the model reproduces well the 
   available experimental data on all quadrupole degrees of freedom, i.e.\
   data on $\beta$ \emph{and} $\gamma$.

 Finally, the model offers a reasonable description of nuclear matter properties,
as predicted by modern ab-initio calculations,  though with a relatively soft equation 
of state for infinite neutron matter.

These qualities render the new model competitive with those based on the 
older BSk effective interactions. While several of the latter (notably HFB-27 based on the BSk27 interaction) 
achieve somewhat lower overall rms deviations on the nuclear masses, the new 
model presents a significant step forward in multiple ways. First, we include for the first time triaxial deformation throughout the model adjustment. 
Second, the coordinate-space representation results in an excellent numerical
precision as compared to an expansion in a limited set of harmonic oscillator
states and results in much smoother trends for the separation energies. 
Third, the model can be used for applications that rely on a coordinate-space
representation without ambiguity.
   
To readjust the parameters of the model, we needed to perform repeated 
three-dimensional calculations in a large single-particle basis. To offset
the inherent computational cost, we presented a new adjustment procedure 
guided by a committee of multilayer neural networks. Each neural network 
provides us with a computationally cheap estimate of observables as a function of 
the parameters, and directs us easily towards its preferred candidate set. 
We improve the training of these networks through active learning on a 
growing set of self-consistent calculations and by pooling the predictions 
of a few hundred committee members, the model adjustment is guided to the 
relevant region of the parameter space. The neural network is only used for the parameter adjustment. Once this procedure is complete, all presented results are derived from a self-consistent EDF calculation.

A particular strength of this approach is its reusability for different 
objective functions. The committee is able to propose a candidate parameter set 
for an arbitrary function of observables, provided the committee is sufficiently
well-trained to reliably estimate them. We exploited this feature to efficiently
search for a compromise between the description of nuclear masses and
realistic pairing strengths.

\subsection{Outlook}
This work opens multiple pathways to even more refined global mass models.
   The first such direction is the lifting of the remaining symmetry restrictions:
   the inclusion of reflection asymmetric and time-reversal breaking nuclear 
   configurations. While the former are expected to play a role for ground states
   in limited regions of the nuclear chart, the latter are relevant 
   for the description of all nuclei with an odd neutron and/or odd proton number. 
   Even though these constitute the majority of all nuclei, the effects of 
   time-reversal breaking have never been studied in a global fashion. 
   These will mainly add a correction to the odd-even staggering of masses that 
   for BSkG1 is governed by pairing correlations.
   Current efforts in this direction are ongoing.
   
We have limited ourselves here to a global discussion of 
   quadrupole deformation, charge radii and nuclear masses.
   A second path concerns the investigation  with the 
   BSkG1 model of other quantities that are known to impact astrophysical applications, and in particular 
   the description of fission. We have shown the impact of triaxiality on the
   nuclear ground state, but this degree of freedom has long been known to 
   affect the height of the first barrier for actinide nuclei~\cite{Larsson72}. 
   Studies of relativistic functionals have shown that it can affect the 
   second barrier systematically~\cite{Lu14}. We suspect this is the case for Skyrme 
   EDFs as well, as hinted in Ref.~\cite{Ryssens19b}, but a global study is still missing. 
   A dedicated study is necessary, whose conclusions can be fed back into the 
   adjustment of future refinements of the mass model.

A third possible improvement concerns our treatment of 
   collectivity. We do not account in any way for shape fluctuations and  
   our approach to rotational collective motion remains highly approximate. 
   Incorporating some degree of configuration mixing into the model, while very 
   demanding,  would allow for improvements on both types of collectivity. 
   A more approachable strategy for nuclear vibration consists of incorporating 
   a simple phenomenological prescription along the lines of Ref.~\cite{Goriely07}. 
   The  rotational correction could be improved by basing it on the Thouless-Valatin
   moments of inertia, which are known to capture the nuclear response to rotation
   more accurately than the Belyaev prescription~\cite{Petrik18}. 
   The calculation of the former necessitates the breaking of time-reversal 
   symmetry that has not been considered here. The extension to the full and 
   systematic calculation of quadrupole and octupole correlation energies 
   through the generate coordinate method represents a difficult path that will have to 
   be taken at some point.    
  
{Future improvements of the mass predictions will also require a better 
 description of nuclei close to the $N=126$ magic number, which are of special relevance 
 to the r-process nucleosynthesis. A possible improvement could be found by 
 following the same path as the BSk series, {\it i.e.}\  by including the 
 density-dependent terms $t_4$ and $t_5$ in the energy density functional. The 
 inclusion of these terms enables the creation of a functional with a low 
 symmetry coefficient $J=30$ MeV that also reproduces a stiff equation of state 
 in infinite neutron matter \cite{Goriely13b}. Guided by ab-initio calculations 
 of the pairing effects in infinite nuclear matter \cite{Cao06}, more realistic 
 forms of the pairing interactions \cite{Chamel10} can also help improve the 
 global coherence of the model.  
  
 All proposed additional ingredients could potentially influence the parameter 
   adjustment with possibly interfering effects and might all require additional
   data to be added as constraints to the objective function. In order to 
   maintain control of the parameter adjustment, they should not be added 
   simultaneously. For this reason and their inherent computational complexity, 
   implementing the ensemble of proposed improvements promises a long road
   ahead, along which the guidance of neural networks in the parameter fit will
   be essential.

\begin{acknowledgements}
We are grateful to Magda Zielinska for extracting the experimental $\gamma$ 
   values from the literature and providing them to us for the preparation
   of  Fig.~\ref{fig:comp_exp_def}, as well as for constructive comments on the manuscript.
This work was supported by the Fonds de la Recherche Scientifique (F.R.S.-FNRS) 
and the Fonds Wetenschappelijk Onderzoek - Vlaanderen (FWO) under the EOS 
Project nr O022818F. The present research benefited from computational resources 
made available on the Tier-1 supercomputer of the F\'ed\'eration 
Wallonie-Bruxelles, infrastructure funded by the Walloon Region under the grant 
agreement nr 1117545. S.G.\ and W.R.\ acknowledge financial support from the 
FNRS (Belgium). W.R. also acknowledges support by the U.S.\ DOE grant 
No. DE-SC0019521. Work by M.B.\ has been supported by the French Agence Nationale 
de la Recherche under grant No.\ 19-CE31-0015-01 (NEWFUN).
\end{acknowledgements}

\appendix

\section{Coupling constants of $E_{\rm Sk}$}
\label{app:couplingconstants}

The Skyrme energy density $\mathcal{E}$ of Eq.~\eqref{eq:Skyrme} is determined by ten 
coupling constants, which are determined by the model parameters $t_{0-3}, x_{0-3}, W_0$ and $W_0'$ as
follows:

\begin{subequations}
\begin{alignat}{2}
C^{\rho\rho}_0               &= \phantom{+}  \tfrac{3}{8}t_0                                  \, ,   \\ 
C^{\rho\rho}_1               &=          -   \tfrac{1}{4}t_0 \left( \tfrac{1}{2} + x_0\right) \, ,   \\
C^{\rho\rho\rho^{\gamma}}_0  &= \phantom{+}  \tfrac{3}{48} t_3  \, ,                                 \\
C^{\rho\rho\rho^{\gamma}}_1  &=          - \tfrac{1}{24} t_3 \left( \tfrac{1}{2} + x_3\right)\, ,  \\
C^{\rho\tau}_0               &= \phantom{+}\tfrac{3}{16} t_1                                   
                                         + \tfrac{1}{4}  t_2 \left( \tfrac{5}{4} + x_2\right)\, ,       \\ 
C^{\rho\tau}_1               &=          - \tfrac{1}{8}  t_1\left( \tfrac{1}{2} + x_1\right)
                                         -               t_2 \left(\tfrac{1}{2} + x_2 \right)  \, ,       \\
C^{\rho \Delta \rho}_0       &=          - \tfrac{9}{64} t_1 + \tfrac{1}{16} t_2 \left(\tfrac{5}{4} + x_2\right) \, , \\ 
C^{\rho \Delta \rho}_1       &= \phantom{+}\tfrac{3}{32}  t_1 (\tfrac{1}{2} + x_1)
                                         + \tfrac{1}{32} t_2  (\tfrac{1}{2} + x_2)  \, ,\\
C^{\rho \nabla \cdot J}_0    &= - \frac{W_0}{2} - \frac{W_0'}{4} \, , \\
C^{\rho \nabla \cdot J}_1    &= - \frac{W_0'}{4} \, .
\end{alignat}
\end{subequations}


\section{Further details on the rotational correction and the MOI}
\label{app:rotational}

The rotational correction, Eq.~\eqref{eq:Erot}, depends on the calculation 
of $\langle \hat{J}_{\mu}^2 \rangle$ and $\mathcal{I}_{\mu}$ for all three
principal axes of the nucleus. Formulas are available in the literature for 
even-even nuclei (see, e.g. \cite{Ryssens15,RingSchuck}). However, 
naively utilizing these expressions in our calculations is problematic 
for two reasons. 

The first is purely technical: the calculation of $\mathcal{I}_{\mu}$
involves a summation over all possible two-quasiparticle excitations in 
the model space, weighted by the inverse of the sum of their 
quasiparticle energies. Unlike any other quantity discussed here, this sum is 
not naturally cut by the single-particle occupation factors.
As our numerical implementation can only represent a fraction of the entire 
quasiparticle spectrum, we have introduced an additional cutoff for the 
rotational correction. We replace the matrix elements of the single-particle 
angular momentum operator $\hat{\jmath}$ that figure into the calculation of 
\emph{both} $\langle \hat{J}^2_{\mu} \rangle$ and $\mathcal{I}_{\mu}$ as
\begin{align}
\langle k | \hat{\jmath}_{\mu} | l\rangle &\rightarrow f_{q, k}^{\rm MOI} f_{q, l}^{\rm MOI}
\langle k | \hat{\jmath}_{\mu} | l\rangle \, , \\
f_{q, k}^{\rm MOI} &= \left[1 + e^{(\epsilon_k - \lambda_q - E_{\rm cut})/\mu_{\rm MOI}}\right]^{-1/4} \, ,
\end{align}
where the cut-off energy $E_{\rm cut}$ is identical to the one used in the pairing channel 
(Eq.~\ref{eq:paircut}), but $\mu_{\rm MOI} = 1$  MeV. 

The second problem affects the calculation of both quantities for 
  odd-$A$ and odd-odd nuclei. For the ground states of even-even nuclei, 
  the expectation value of  $\hat{J}_{\mu}^2$ and the Belyaev MOI are purely collective 
  in nature, {\it i.e.}\ non-zero values are generated by many nucleons as a result of 
  the nuclear deformation. This picture is modified significantly by the 
  presence of blocked quasiparticles, whose individual contributions to 
  both $\hat{J}_{\mu}^2$ and $\mathcal{I}_B$ are generally sizeable and 
  cannot be considered as collective. Furthermore, the Belyaev MOI
  is fundamentally a quantity obtained from second-order perturbation theory
  of an HFB minimum. While its calculation can be generalized to include the 
  possibility of blocked quasiparticles along the lines of Ref.~\cite{Alhassid}, 
  the validity of such an approach can be questioned. As a purely practical 
  recipe to sidestep these issues, we calculate both $\hat{J}_{\mu}^2$ and 
  $\mathcal{I}_B$ for odd-$A$ and odd-odd nuclei by omitting the contributions
  from all blocked quasiparticles, mirroring the approach of Ref.~\cite{Koh16}.

Finally, we comment on the comparison of calculated (Belyaev) MOI
 with experimental data, as we do in Fig.~\ref{fig:MOI}. 
  For axial configurations, the Belyaev MOI  along the 
  symmetry axis vanishes, while the two remaining values are equal; comparison
  to experiment is then straightforward. For triaxial nuclear configurations, 
  we obtain however three non-zero, distinct values for the MOI. In those
  cases, we have chosen systematically the largest among the three values as
  the one to be compared to experiment. We have made this rather ad-hoc choice 
  motivated by a naive semi-classical model of rotation, where the largest 
  MOI produces the lowest-lying rotational excitations. As the 
  experimental data is extracted from the excitation energy of 
  the first $2^+$ state in rotational nuclei, this seems to be the most 
  appropriate choice. 

\section{Explanation of the supplementary material}

We provide as supplementary material the file \newline
\textsf{Mass\_Table\_BSkG1.dat}, 
which contains the calculated ground state properties of all nuclei with 
$ 8 \leq Z \leq 110 $ lying between the proton and neutron drip lines. Its content is summarized and 
explained in Tab.~\ref{tab:suppl}. A few additional remarks are in order:
\begin{itemize}
\item \textbf{Column 11/12}: A unique definition of the pairing gap exists only
                             for HFB calculations with schematic interactions. 
                             To extract some information on the overall importance
                             of the pairing correlations for a given nucleus, 
                             we use the $uv$-weigthed average pairing gaps 
                             $\langle \Delta \rangle_{n/p} $ of Ref.~\cite{Bender00}.
\item \textbf{Column 16}: we report only the largest MOI among all
      three directions, {\it i.e.}\ the file contains \newline $\max_{\mu = x,y,z} \{ \mathcal{I}_{\mu} \}$. 
\item \textbf{Column 17/18}: as discussed in Sec.~\ref{sec:oddnuclei}, we construct auxiliary states for odd-A and odd-odd nuclei through a self-consistent blocking procedure. To make these calculations reproducible, we provide for such nuclei the parity quantum number of the blocked quasiparticles.
\end{itemize}

\begin{table*}[]
\begin{tabular}{llll}
\hline
\hline
Column &  Quantity & Units & Explanation \\
\hline
1 & Z & $-$ & Proton number\\
2 & N & $-$ & Neutron number\\
3 & $M_{\rm exp}$ & MeV  & Experimental atomic mass excess \\
4 & $M_{\rm th}$ & MeV   & BSkG1  atomic mass excess \\
5 & $\Delta M$ & MeV      & $M_{\rm exp} - M_{\rm th}$\\
6 & $E_{\rm tot}$ & MeV   & Total binding energy, Eq.~\eqref{eq:Etot}. \\
7 & $\beta_{20}$ & $-$    & Deformation, Eq.~\eqref{eq:betalm}.\\
8 & $\beta_{22}$ & $-$    & Deformation, Eq.~\eqref{eq:betalm}.\\
9 & $\beta$  & $-$    & Deformation, Eq.~\eqref{eq:beta}\\
10 & $E_{\rm rot}$ & MeV  & Rotational correction, Eq.~\eqref{eq:Erot}. \\
11 & $\langle \Delta \rangle_n$ & MeV & Average neutron gap\\
12 & $\langle \Delta \rangle_p$ & MeV & Average proton gap\\
13 & $r_{\rm BSkG1}$ & fm  & Calculated rms charge radius\\
14 & $r_{\rm exp}$  & fm  & Experimental rms charge radius\\
15 &$\Delta r$ & fm   &  $r_{\rm exp}- r_{\rm BSkG1}$  \\
16 & $\mathcal{I}_{\rm MOI}$ & $\hbar^2$ MeV$^{-1}$ & Calculated MOI.\\
17 & par(p) & $-$ & Parity of protons qp. excitation\\
18 & par(n) & $-$ & Parity of neutrons qp. excitation\\
\hline 
\hline
\end{tabular}
\caption{Contents of the \textsf{Mass\_Table\_BSkG1.dat} file.}
\label{tab:suppl}
\end{table*}


\newpage{\pagestyle{empty}\cleardoublepage}

\end{document}